\documentclass[11pt,a4paper]{article}

\usepackage{amsmath,amsthm,amssymb}
\usepackage{graphics,graphicx}
\usepackage{epsfig}
\usepackage{multicol}
\usepackage{color}
\makeatletter
\@addtoreset{equation}{section}
\makeatother

\def\be{\begin{equation}}
\def\ee{\end{equation}}
\def\bea{\begin{eqnarray}}
\def\eea{\end{eqnarray}}

\begin{document}

\begin{center}
\LARGE{\bf Dynamical behavior and Jacobi stability analysis of wound strings}
\end{center}

\begin{center}
\large{\bf Matthew J. Lake} ${}^{a,b,}$\footnote{matthewj@nu.ac.th} and \large{\bf Tiberiu Harko} ${}^{c,d,}$\footnote{t.harko@ucl.ac.uk}
\end{center}
\begin{center}

\emph{ ${}^a$ The Institute for Fundamental Study, ``The Tah Poe Academia Institute", \\ Naresuan University, Phitsanulok 65000, Thailand\\}
\emph{ ${}^b$ Thailand Center of Excellence in Physics, Ministry of Education, Bangkok 10400, Thailand\\}
\emph{ ${}^c$ Department of Physics, Babes-Bolyai University, Kogalniceanu Street,\\
Cluj-Napoca 400084, Romania \\}
\emph{ ${}^d$ Department of Mathematics, University College London, \\ Gower Street, London WC1E 6BT, United Kingdom \\}
\vspace{0.1cm}
\end{center}

\begin{abstract}

We numerically solve the equations of motion (EOM) for two models of circular cosmic string loops with windings in a simply connected internal space. Since the windings cannot be topologically stabilized, stability must be achieved (if at all) dynamically. As toy models for realistic compactifications, we consider windings on a small section of $\mathbb{R}^2$, which is valid as an approximation to any simply connected internal manifold if the winding radius is sufficiently small, and windings on an $S^2$ of constant radius $\mathcal{R}$. We then use Kosambi-Cartan-Chern (KCC) theory to analyze the Jacobi stability of the string equations and determine bounds on the physical parameters that ensure dynamical stability of the windings. We find that, for the same initial conditions, the curvature and topology of the internal space have nontrivial effects on the microscopic behavior of the string in the higher dimensions, but that the macroscopic behavior is remarkably insensitive to the details of the motion in the compact space. This suggests that higher-dimensional signatures may be extremely difficult to detect in the effective $(3+1)$-dimensional dynamics of strings compactified on an internal space, even if configurations with nontrivial windings persist over long time periods.

\end{abstract}

\section{Introduction} \label{S1}

The interpretation of physical theories in geometric terms has a long history in theoretical physics, and such geometrodynamical approaches have been used for a long time to model gravitational phenomena in the framework of general relativity, string theory, and in many other applications (see, for example \cite{Frankel:1997ec,Nakahara:1990th}). However, geometric methods can be employed in an equally successful way to formulate a general geometric theory of dynamical systems, or, more precisely, of systems of second order ordinary differential equations. For example, a geometric interpretation of classical mechanics can be obtained in this way \cite{Pet1}-\cite{Pet4}. 

One very powerful approach for studying the properties of systems of differential equations using geometrical methods is Kosambi-Cartan-Chern (KCC) theory, which was initiated in the pioneering works of Kosambi \cite{Ko33}, Cartan \cite{Ca33} and Chern \cite{Ch39}, respectively. The KCC theory was inspired by, and proposed and developed in, the geometric framework arising in the study of Finsler spaces \cite{An00,Mo,ChernAMS}. The theory is based on the fundamental assumption that there is a one to one map between a \textit{system of ordinary differential equations, which is globally defined, and the coefficients of a semispray} \cite{An00}. (For a recent review of the KCC theory see \cite{rev}.)

In the geometrical description of dynamical systems proposed by KCC theory, two geometric quantities are associated to each system of second order differential equations: a non-linear connection, and a Berwald type connection, respectively. (These are discussed in detail in Sect. \ref{S4}.) With the use of these two connections, five geometric invariants of the system are then constructed. From a physical perspective, the most important is the second invariant $P_j^i$, called the curvature deviation tensor. Its importance lies in the fact that the sign of the real parts of the eigenvalues of $P_j^i$ determine the Jacobi stability (or instability) of the solutions of the system of differential equations to which $P_j^i$ corresponds \cite{An00,rev,Sa05,Sa05a}. The solutions are said to be Jacobi stable if the trajectories of the dynamical system converge in the vicinity of a critical point of the system, and Jacobi unstable if they diverge.

The system of differential equations describing the deviations of the whole trajectory of a dynamical system, with respect to perturbed trajectories, are introduced geometrically in KCC theory via the second KCC invariant \cite{An00, rev}, and KCC theory has been extensively applied in studies of the stability of different physical, engineering, biological, and chemical systems \cite{Sa05}-\cite{Ha5}. In the present work, we apply KCC theory to the study of fundamental, or ``$F$-string" loops, with windings in the compact internal space predicted by string theory \cite{Zwi09,Polchinski:1998rq,Polchinski:1998rr,Green:2012oqa}. However, our analysis applies equally well to an species of cosmic strings, represented in the ``wire approximation" \cite{ViSh00,And02}, in any higher-dimensional scenario.

We begin by considering circular loops of string, in $(3+1)$ dimensions, with windings on a small section of $\mathbb{R}^2$ representing the higher-dimensional space. Though $\mathbb{R}^2$ is \emph{not} compacfified, the higher-dimensional part of the metric is intended as an approximation to \emph{any} genuinely compact internal manifold, when the radius of the string windings is sufficiently small. In this case, both the overall topology of the internal space and the local curvature can be neglected. By comparing this simple example (Model I) with the ``simplest", topologically nontrivial, simply connected internal space, an $S^2$ (Model II), we aim to highlight the role that the topology and curvature of the internal space play in the both the higher-dimensional dynamics and the effective $(3+1)$-dimensional dynamics of the string.

Generally, we find that the dynamics in the internal space are extremely sensitive to the initial conditions and values of the model parameters, including the winding number and initial velocities of the string in the compact directions, and exhibit a rich variety of qualitatively different periodic evolutions, often with several periodicities superimposed in a nontrivial way. In contrast, the $(3+1)$-dimensional dynamics are determined primarily by the initial loop radius $\rho_0$ and initial velocity of the string in the Minkowski directions (as expected), together with the value of an additional parameter, $\Omega_0^2 \in (0,1)$, which represents the initial fraction of the total string length lying in the large dimensions. The $(3+1)$-dimensional loop radius exhibits an extremely uniform oscillatory signature, with a single oscillation period which is determined by these parameters. Therefore, it makes very little difference how rapidly the string oscillates in the internal space, or whether its oscillations contain a single or multiple periodicities $-$ in all cases, the effect on the macroscopic string dynamics remains negligible over very large time scales.

It is intuitively straightforward to see why this should be the case \emph{if} the oscillations in the compact manifold remain small. Bearing this in mind, perhaps the most important difference between the $\mathbb{R}^2$ and $S^2$ examples considered here is that the oscillations in the compact space remain regular (do not increase with time) in the former, but grow increasingly rapidly in the latter. Nonetheless, they remain small over long time periods, compared to the oscillations of the loop in the Minkowski directions, and have a negligible effect on the macroscopic string dynamics, {\it except} via their contribution to $\Omega_0^2$. To clarify, the motion of the string in the compact space {\it does} affect its macroscopic motion, but mainly through the contribution to the ``bulk" parameter $\Omega_0^2$. In pioneering work, Nielsen showed that the motion of windings in a compact internal space gives rise to an effective world-sheet current, $j$, from a $(3+1)$-dimensional perspective, via dimensional reduction \cite{Nielsen:1979zf,Nielsen:1987fy}. Hence, in our model, $\Omega_0^2$ is related to the initial integrated current, which, by the circular symmetry of the string loop, is proportional to the instantaneous current. (In fact, it is straightforward to show that, for circular string loops, $j(t_0) \propto (1-\Omega_0^2)/\Omega_0^2$ \cite{Yamauchi:2014ita}.) What is remarkable is that the subsequent evolution of the string in the Minkowski directions is largely independent of the detailed dynamics in the compact space. This indicates that, in many cases, different higher-dimensional trajectories are {\it degenerate} from a $(3+1)$-dimensional perspective. Integrating out over the compact directions (performing dimensional reduction), they give rise to the same time-evolution for $j(t)$ and, hence, $\rho(t)$.

Unfortunately, due to computational restrictions, we were unable to solve the string EOM over a very large number of cycles in the $(3+1)$-dimensional loops radius (our time scale was limited to $\sim \mathcal{O}(10)-\mathcal{O}(10^2)$ complete oscillations). Therefore, we have not been able to \emph{entirely} rule out the possibility of a late onset of instability in the $(3+1)$-dimensional dynamics due to higher dimensional effects. That said, there is at present no numerical evidence for such an effect. In addition, in the Jacobi stability analysis of the string EOM, we find no unstable critical points in either the $\mathbb{R}^2$ or $S^2$ models. These factors combined strongly suggest that, at least in these examples, the motion of the string in the (visible) large dimensions yields no observable signatures of higher-dimensional physics. One further limitation of the present analysis is that it is entirely classical, though quantum effects may be expected to influence the motion of the string in the compact space. We comment briefly on this in the Discussion, Sect. \ref{S7}. For now, we note that, by \cite{Nielsen:1979zf,Nielsen:1987fy}, classical treatment of the higher-dimensional string motion is equivalent to classical treatment of the superconducting string current, which remains approximately valid for $j$ less than the threshold value, $j_{\max} \propto R^{-1}$, where $R$ is the string width. This marks the onset of a quantum instability, which causes the superconducting string to decay \cite{Witten85}.

Though, in principle, it is possible that more complex internal manifolds may yield significantly different results (due to the strongly nonlinear nature of the EOM, this is somewhat difficult to predict), our results suggest that the effective $(3+1)$-dimensional dynamics of higher-dimensional cosmic strings could, \emph{in general}, be remarkably insensitive to their dynamics in the compact internal space. Thus, even if higher-dimensional strings exist, and even if the detection of either gravitational waves or gauge particle emission from strings is one day confirmed, it seems unlikely that the higher-dimensional nature of the string motion will leave any significant imprint on these observable signatures. We contend that, though is primarily a ``negative result", such results are necessary if we one day hope to be able to distinguish genuine signatures of higher-dimensional physics, which \emph{may} be present in observations of high energy astrophysical objects, from signatures predicted by conventional $(3+1)$-dimensional models. We need to know what, and what not, to look for.

The structure of this paper is as follows. In Sect. \ref{S2}, we review the action, EOM, energy-momentum tensor, and boundary conditions for the $F$-string. In Sect. \ref{S3}, we introduce the wound string ansatz and determine the exact form of the EOM for each model. We then solve the EOM numerically for a variety of initial conditions and determine the critical points of the system. (The models are considered sequentially, first Model I, then Model II.) In Sect. \ref{S4}, we give a detailed review of the mathematical and physical basis of KCC theory, and the theory of Jacobi stability (or instability) of a system of differential equations. In Sects. \ref{S5} and \ref{S6}, the KCC formalism is applied to the critical points of Models I$-$II, respectively, which were identified in Sect. \ref{S3}. The critical points of both systems are found to be Jacobi stable, in agreement with our previous numerical results. A summary of both the analytic and numerical results obtained in our analysis is given in Sect. \ref{S7}, and prospects for future work are briefly discussed.

\section{The $F$-string action, EOM, and energy-momentum tensor} \label{S2}
In the absence of world-sheet fluxes, the string is governed by the Nambu-Goto action \cite{Go71,Na77}
\begin{eqnarray} \label{Act_2.1}
S = -\mathcal{T}\int {\rm d}^2\zeta \sqrt{-\gamma},
\end{eqnarray}
where $\gamma$ is the determinant of the induced metric on the world-sheet
\begin{eqnarray} \label{IndMet_2.1}
\gamma_{ab}(\zeta) = g_{IJ}\left(X\right) \frac{\partial X^{I}}{\partial \zeta^{a}}\frac{\partial X^{J}}{\partial \zeta^{b}}.
\end{eqnarray}
Here, $I,J \in \left\{0,1,2,3, \dots  d\right\}$ label the space-time embedding coordinates and $a,b \in \left\{0,1\right\}$ where $\zeta^{0}=\tau$ and $\zeta^{1}=\sigma$ denote the time-like and space-like world-sheet coordinates, respectively. The intrinsic string tension is $\mathcal{T} = 1/(2\pi \alpha')$, where $\alpha'$ is the Regge slope parameter, which is related to the fundamental string scale via $l_{st} = \sqrt{\alpha'}$, in natural units, $\hbar = 1$, $c = 1$ \cite{Zwi09}.

Using the identity $\delta(-\gamma) = (-\gamma)\gamma^{ab}\delta\gamma_{ab}$, variation of the action yields the covariant string EOM \cite{And02}
\begin{eqnarray} \label{AltEL_2.1}
\frac{\partial}{\partial \zeta^{a}}\left(\sqrt{-\gamma}\gamma^{ab}g_{IJ}(X)\partial_{b}X^{J}\right) - \frac{1}{2}\sqrt{-\gamma}\gamma^{cd}\frac{\partial g_{KL}(X)}{\partial X^{I}}\partial_{c}X^{K}\partial_{d}X^{L}= 0,
\end{eqnarray}
plus a boundary term
\begin{eqnarray} \label{BoundTerm_2.1}
\left[\mathcal{P}^{\sigma}{}_I(\tau,\sigma) \delta X^I  \right]_{0}^{\sigma_f}  = 0,
\end{eqnarray}
where
\begin{eqnarray} \label{CanMom_2.1}
\mathcal{P}^{a}{}_I(\tau,\sigma) = \frac{\partial(\mathcal{L}\sqrt{-\gamma})}{\partial (\partial_a X^I)}
\end{eqnarray}
is the canonical momentum of $X^{I}$ with respect to $\zeta^{a}$. Without loss of generality, we may assume that the space-like world-sheet parameter varies over the range $\sigma \in [0,\sigma_f]$. To satisfy the boundary term, we may impose Dirichlet, Neumann or periodic boundary conditions:
\begin{eqnarray} \label{Dir_2.1}
X^{I}(\tau,0) = const., \ \ \ X^{I}(\tau,\sigma_f) = const.,
\end{eqnarray}
\begin{eqnarray} \label{Neu_2.1}
\mathcal{P}^{\sigma}{}_I(\tau,0) = \mathcal{P}^{\sigma}{}_I(\tau,\sigma_f) = 0,
\end{eqnarray}
\begin{eqnarray} \label{Per_2.1}
X^{I}(\tau,\sigma) = X^{I}(\tau,\sigma + m\sigma_f), \ \ \ m \in \mathbb{Z},
\end{eqnarray}
respectively, where the imposition of Neumann boundary conditions implies that the string end points move at the speed of light \cite{Zwi09}.

Variation of the Nambu-Goto action with respect to the space-time metric $g_{IJ}(x)$, where $x^{I}$ denotes a space-time background coordinate, gives \cite{ViSh00,And02}
\begin{eqnarray} \label{T_mu_nu}
T^{IJ} = \frac{1}{\sqrt{-g}}\mathcal{T}\int \sqrt{-\gamma} \gamma^{ab}\partial_{a}X^{I}\partial_{b}X^{J} \delta^D(x-X){\rm d}\tau {\rm d}\sigma.
\end{eqnarray}
and  it is straightforward to verify that, for a system of embedding coordinates in which $X^{0} \propto x^{0} \propto \zeta^{0}$, $X^{1} \propto x^{1} \propto \zeta^{1}$, the covariant EOM (\ref{AltEL_2.1}) are in one-to-one correspondence with the conservation equations \cite{Dir75}
\begin{eqnarray} \label{ConsLawAlt_2.3}
\nabla_{J}T^{J}{}_{I}\sqrt{-g} = \frac{\partial}{\partial x^{J}}\left(T^{J}{}_{I}\sqrt{-g}\right)  - \frac{1}{2} \frac{\partial g_{AB}(x)}{\partial x^{I}}T^{AB}\sqrt{-g} = 0.
\end{eqnarray}
For the models considered in this work, we choose background coordinates, an embedding, and a world-sheet parameterization that ensures this correspondence holds, so that the string EOM are direct expressions of energy-momentum conservation.

\section{Numerical solution of the string equations} \label{S3}

In the section, we define the ansatz for the circular wound string loop in the static gauge ($X^{0} \propto \tau$), using cylindrical polar coordinates. The EOM for strings with windings on $\mathbb{R}^2$ and $S^2$ internal spaces (Model I and Model II, respectively) are determined and solved numerically, and the critical points of each model are determined analytically.

\subsection{Model I: numerical solutions of the EOM for windings on $\mathbb{R}^2$} \label{S3.1}

In our first model, we assume that the background geometry is (effectively) described by the metric with line element
\begin{equation}
ds^2 = a^2({\rm d}t^2 - {\rm d}\rho^2 - \rho^2{\rm d}\sigma^2 - {\rm d}z^2) - {\rm d}R^2 - R^2{\rm d}\phi^2,
\end{equation}
where $a^2 \in (0,1]$ is a phenomenological ``warp factor", which accounts for the back reaction on the large dimensions when the internal dimensions are flux-compactified, as expected in string theory \cite{Denef:2007pq}. Here $R$ is the radial coordinate and $\phi$ is the angular coordinate in the higher-dimensional plane. The ansatz for the circularly symmetric wound string loop is
\begin{equation}
X^{I} = (t=\xi \tau, \ \rho(t), \ \sigma, z=0, \ R(t), \ \phi(t,\sigma) = m\sigma + f(t))
\end{equation}
where $\xi$ is a constant with dimensions of length and $m \in \mathbb{Z}$, so that
\begin{equation}
(-\gamma) = \xi^2\left[(a^2(1-\dot{\rho}^2) - \dot{R}^2)(a^2\rho^2 + m^2R^2) - a^2\rho^2R^2\dot{\phi}^2\right],
\end{equation}
where a dot represents differentiation with respect to $t$.

The string EOM are
\begin{equation}
\frac{{\rm d}}{{\rm d}t}\left( \frac{a^{2}\rho ^{2}+m^{2}R^{2}}{\sqrt{-\gamma }}\right)
=0,  \label{1}
\end{equation}

\begin{equation}
\frac{{\rm d}}{{\rm d}t}\left( \frac{\rho ^{2}R^{2}\dot{\phi}}{\sqrt{-\gamma }}\right)
=0,  \label{2}
\end{equation}

\begin{equation}
\left(a^{2}\rho ^{2}+m^{2}R^{2}\right) \ddot{\rho}+\rho \left[ a^{2}\left(1-\dot{\rho}^{2}\right) -\dot{R}^{2}-R^{2}\dot{\phi}^{2}\right] =0,
\label{3}
\end{equation}

\begin{equation}
\left(a^{2}\rho ^{2}+m^{2}R^{2}\right) \ddot{R} + R\left\{m^{2}\left[a^{2}\left(1-\dot{\rho}^{2}\right) -\dot{R}^{2}\right] - a^{2}\rho ^{2}\dot{\phi}^{2}\right\} =0.  \label{4}
\end{equation}
By combining Eqs. (\ref{1}) and (\ref{2}) we obtain
\begin{equation}
\dot{\phi}=C\frac{a^{2}\rho ^{2}+m^{2}R^{2}}{\rho ^{2}R^{2}},  \label{5}
\end{equation}
where $C$ is an arbitrary constant of integration. Physically, it is equal to the ratio of the conserved momentum in the $\phi$-direction to the string energy,
\begin{equation} \label{C = l/E}
C = l/E,
\end{equation}
where
\begin{equation} \label{E-l}
E = 2\pi\mathcal{T} \frac{a^{2}\rho ^{2}+m^{2}R^{2}}{\sqrt{-\gamma }}, \ \ \ \ l = 2\pi\mathcal{T} \frac{\rho ^{2}R^{2}\dot{\phi}}{\sqrt{-\gamma }}.
\end{equation}
A useful model parameter is
\begin{equation}
\Omega^2 = \frac{a^{2}\rho ^{2}}{a^{2}\rho ^{2}+m^{2}R^{2}},
\end{equation}
which represents the (time-dependent) fraction of the total string length in the large dimensions \cite{Yamauchi:2014ita,LaWa10,LaYo12} and, in this notation, Eq. (\ref{5}) may be rewritten as
\begin{equation}
\dot{\phi} = \frac{a^2}{R^2}\Omega^{-2}C.
\end{equation}

For $R = const.$, Eqs. (\ref{3}) and (\ref{4}) reduce to
\begin{equation} \label{R=const_EOM1}
(1-\dot{\rho}^2)(2\Omega^2-1) + \rho\ddot{\rho} = 0,
\end{equation}
and
\begin{equation}  \label{R=const_EOM2}
\dot{\phi} = \pm \frac{\sqrt{1-\dot{\rho}^2}}{\rho}m,
\end{equation}
respectively. In this case, $(-\gamma) = \xi^2a^2(1-\dot{\rho}^2)\rho^2$ at the level of the EOM, so that
\begin{equation}
E = 2\pi\mathcal{T} \frac{a^2\rho}{\sqrt{1-\dot{\rho}^2}}\Omega^{-2}, \ \ \ \ l = \pm 2\pi\mathcal{T}mR^2,
\end{equation}
and
\begin{equation}
C = \pm \frac{R^2}{a^2}\frac{m}{\rho}\sqrt{1-\dot{\rho}^2}\Omega^2
= \pm \frac{R}{a}\sqrt{1-\dot{\rho}^2} \Omega \sqrt{1-\Omega^2}.
\end{equation}
Eq. (\ref{R=const_EOM1}) can be solved analytically and the general solution is of the form
\begin{equation}
\rho(t) = \mathcal{A}\sqrt{1+\mathcal{B}\sin^2(\mathcal{C}t + \mathcal{D})},
\end{equation}
where the constants $\left\{\mathcal{A},\mathcal{B},\mathcal{C},\mathcal{D}\right\}$ depend on the initial conditions. The general expressions are complex, but for $\dot{\rho}_0 \equiv \dot{\rho}(t_0)=0$, where $t_0$ is the initial time (assumed to be the epoch of loop formation), we have \cite{LaYo12}
\begin{equation}  \label{R=const_EOM3}
\rho(t) = \rho_0\sqrt{1+\left(\frac{1-2\Omega^2_0}{\Omega^4_0}\right)\sin^2\left(\frac{\Omega^2_0}{\rho_0}(t-t_0)\right)},
\end{equation}
where the subscript ``$0$" indicates the quantities evaluated at $t_0$. The functions $\rho(t)$, $\Omega(t)$ and $\lambda(t) = \rho(t)/m$, the distance between windings (or ``wavelength") in $(3+1)$-dimensional space, oscillate between the critical values
\begin{eqnarray}  
\rho_{c1} &=& \rho_0, \ \ \ \rho_{c2} = \left(\frac{1-\Omega^2_0}{\Omega^2_0}\right)\rho_0,
\nonumber\\
\Omega_{c1} &=& \Omega_0, \ \ \ \ \Omega_{c2} = \sqrt{1-\Omega^2_0},
\nonumber\\
\lambda_{c1} &=& 2\pi \frac{\Omega_0}{\sqrt{1-\Omega^2_0}}R, \ \ \ \lambda_{c2} = 2\pi \frac{\sqrt{1-\Omega^2_0}}{\Omega_0} R.
\end{eqnarray}
This scenario is possible, dynamically, for fine-tuned initial conditions, even if the winding radius $R$ is not topologically stabilized. For $R = const.$, the string is effectively tensionless, from a $(3+1)$-dimensional perspective, when $\Omega^2 = 1/2$, giving $C = \pm (1/2)R/a$, $\dot{\phi} = \pm a/R$ and $\rho = const.$ \cite{Yamauchi:2014ita,LaYo12}.

However, for general initial conditions, the system of equations (\ref{3})-(\ref{5}) cannot be solved analytically, and the winding radius $R$ undergoes dynamical evolution. In order to solve the system numerically, we first substitute Eq. (\ref{5}) into Eqs. (\ref{3}) and (\ref{4}), giving the following evolution equations for $\rho $ and $R$:
\begin{equation}
\ddot{\rho}+\frac{\rho }{a^{2}\rho^{2} + m^{2}R^{2}}\left[ a^{2}\left(1-\dot{\rho}^{2}\right)
- \dot{R}^{2}-C^{2}\frac{\left(a^{2}\rho^{2}+m^{2}R^{2}\right)^{2}}{\rho ^{4}R^{2}}\right] =0,  \label{6}
\end{equation}
\begin{equation}
\ddot{R}+\frac{R}{a^{2}\rho^{2} + m^{2}R^{2}}\left[m^{2}\left[a^{2}\left(1-\dot{\rho}^{2}\right) -\dot{R}^{2}\right]
- a^{2}C^{2}\frac{\left(a^{2}\rho ^{2}+m^{2}R^{2}\right)^{2}}{\rho ^{2}R^{4}}\right] =0.  \label{7}
\end{equation}

\begin{figure}[h] 
\centering
\includegraphics[width=8cm]{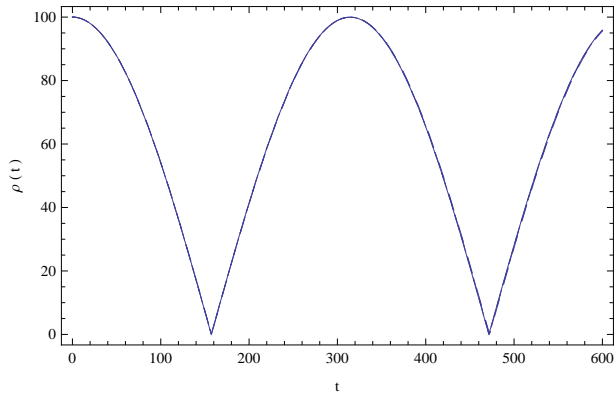}
\caption{Time variation of $\protect\rho(t)$ in Model I for different values of $m$: $m=1$ ({\it solid curve}), $m=2$ ({\it dotted curve}), $m=3$ ({\it short dashed curve}) and $m=4$ ({\it long dashed curve}). The macroscopic motion of the string is extremely insensitive to the value of the winding number and the four curves are practically indistinguishable over the time period shown.}
\label{f1}
\end{figure}

\begin{figure}[h] 
\centering
\includegraphics[width=8cm]{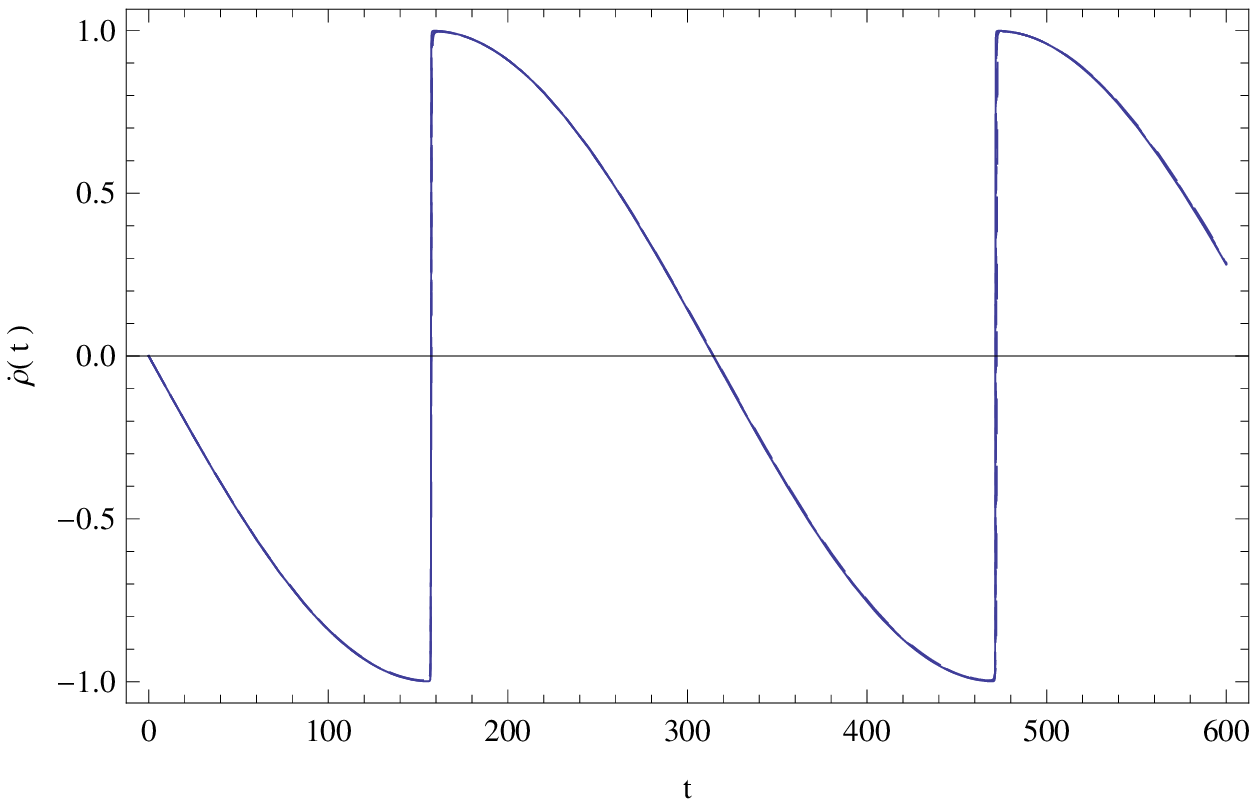}
\caption{Time variation of $\dot{\protect\rho}(t)$ in Model I for different values of $m$: $m=1$ ({\it solid curve}), $m=2$ ({\it dotted curve}), $m=3$ ({\it short dashed curve}) and $m=4$ ({\it long dashed curve}). As in Fig. \ref{f1}, the four curves are practically indistinguishable due to the insensitivity of the macroscopic string motion to the value of the winding number.}
\label{f2}
\end{figure}

\begin{figure}[h] 
\centering
\includegraphics[width=8cm]{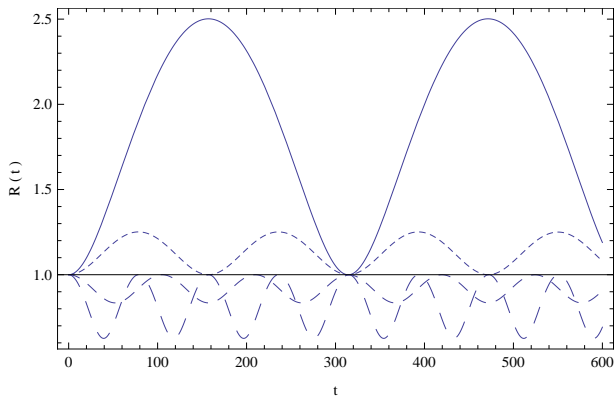}
\caption{Time variation of $R(t)$ in Model I for different values of $m$: $m=1$ ({\it solid curve}), $m=2$ ({\it dotted curve}), $m=3$ ({\it short dashed curve}) and $m=4$ ({\it long dashed curve}). The dynamical evolution of the higher-dimensional winding radius is extremely sensitive to the total number of windings contained in the loop.}
\label{f3}
\end{figure}

\begin{figure}[h] 
\centering
\includegraphics[width=8cm]{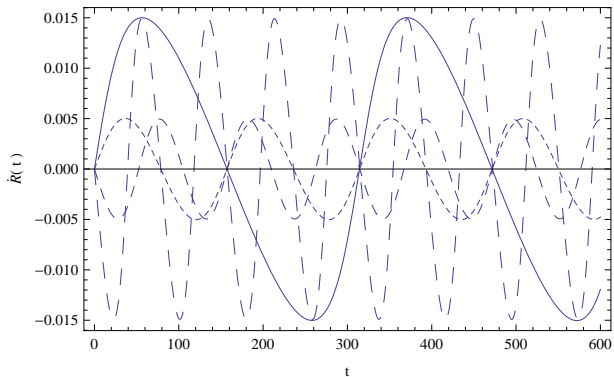}
\caption{Time variation of $\dot{R}(t)$ in Model I for different values of $m$: $m=1$ ({\it solid curve}), $m=2$ ({\it dotted curve}), $m=3$ ({\it short dashed curve}) and $m=4$ ({\it long dashed curve}). As in Fig. \ref{f3}, the sensitivity of the time evolution of the winding radius to the value of the winding number is clearly apparent, and all four curves are easily distinguished.}
\label{f4}
\end{figure}

\begin{figure}[h] 
\centering
\includegraphics[width=8cm]{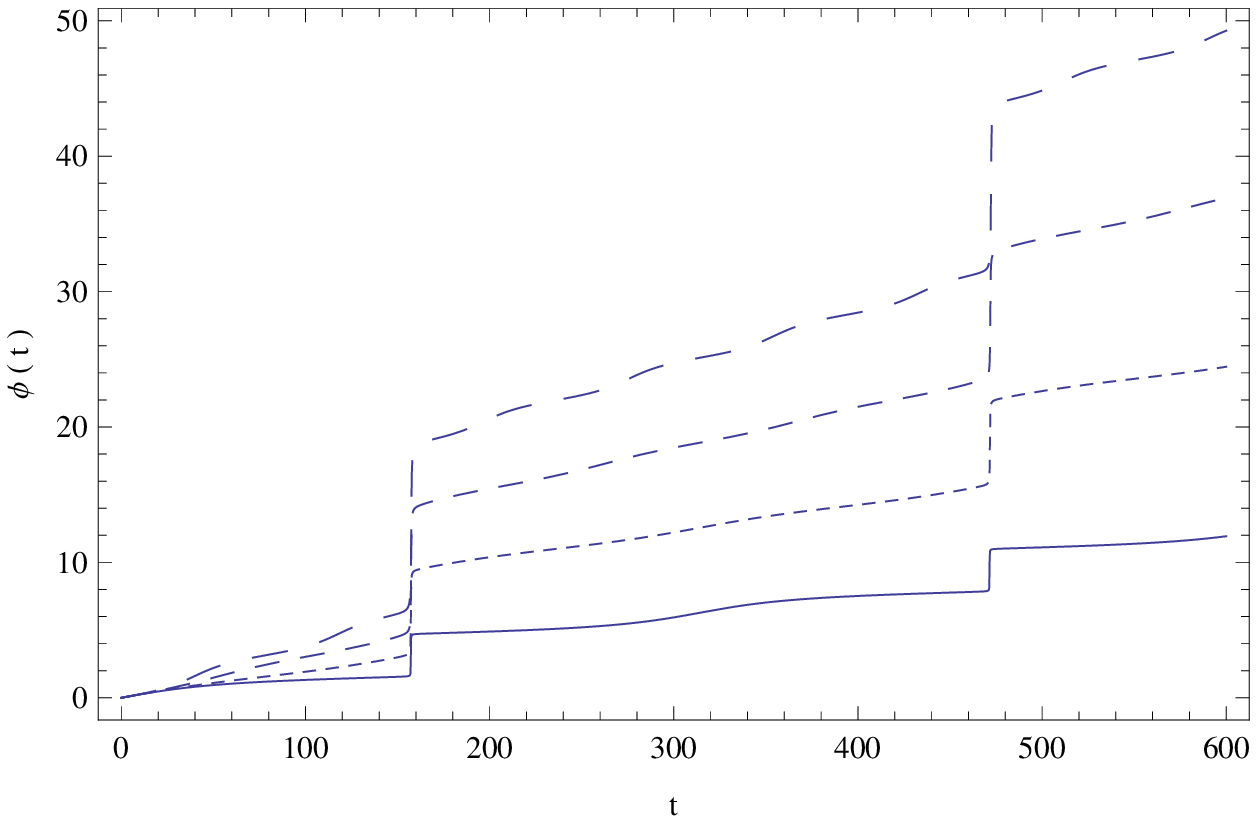}
\caption{Time variation of $\protect\phi(t)$ in Model I for different values of $m$: $m=1$ ({\it solid curve}), $m=2$ ({\it dotted curve}), $m=3$ ({\it short dashed curve}) and $m=4$ ({\it dashed curve}). Though the period of the higher-dimensional angular coordinate matches that of the macroscopic loops radius $\rho(t)$, {\it not} the winding radius $R(t)$, its evolution is sensitive to the value of the winding number, like $R(t)$ (but {\it unlike} $\rho(t)$).}
\label{f5}
\end{figure}

The system of Eqs. (\ref{6}) and (\ref{7}) must be integrated with the initial conditions $\rho \left( 0\right) =\rho _{0}$, $\dot{\rho}(0)=\dot{\rho}_{0}$, $R\left( 0\right) =R_{0}$, $\dot{R}(0)=\dot{R}_{0}$, and $\phi \left( 0\right) =\phi _{0}$, respectively. The time variations of $\rho$, $\dot{\rho}$, $R$, $\dot{R}$ and $\phi $ are represented, for $a=1/\sqrt{2}$, $C=0.05$, and for different values of $m$, in Figs. \ref{f1}, \ref{f2}, \ref{f3}, \ref{f4} and \ref{f5}. The initial values used to integrate Eqs. (\ref{6}) and (\ref{7}) are $\rho (0)=100$, $\dot{\rho}(0)=0$, $R(0)=1$, $\dot{R}(0)=0$ and $\phi (0)=0$.

Note that in Figs. \ref{f1} and \ref{f3}, both the string radius in the large dimensions and the winding radius in the higher-dimensional space do not go to zero due to the conservation of angular momentum. The transition between the expanding and contracting phases is extremely sharp for $\rho$, compared to the overall time period of the oscillation, whereas it is smooth for $R$. Figure \ref{f1} shows that, for these initial conditions, the macroscopic behavior of the string (i.e. the time-evolution of $\rho$) is extremely insensitive to its motion in the compact space, though it may be verified that this is true for a wide range of initial data as well as for much higher values of $m$. On the time scale shown, containing roughly two complete oscillations in $\rho(t)$, the curves corresponding to $m \in \left\{1,2,3,4\right\}$ in Figs. \ref{f1} and \ref{f2} are practically indistinguishable.

As we will see in Sect. \ref{S3.3}, this is true not only for compactifications in which the topology and curvature of the internal space can be safely ignored (i.e. when the winding radius is extremely small), but also in the more general $S^2$ scenario, when the topological and geometric properties of the internal space \emph{do} significantly affect the behavior of the string in the compact dimensions. Interestingly, the behavior of the $\phi$ coordinate, Fig. \ref{f5}, is like that of $\rho$, not $R$, in that the string experiences strong acceleration during the sharp transition between expanding and contracting phases, but evolves relatively smoothly in between. More importantly, the periodicities of $\rho$ and $\phi$ are equal, whereas the period of $R$ may be less than or greater than these, depending on the value of the winding number.

For our chosen initial conditions, the four cases plotted in Figs. \ref{f1}, \ref{f2}, \ref{f3}, \ref{f4} and \ref{f5}, with winding numbers $m \in \left\{1,2,3,4\right\}$, correspond to the following initial fractions of the string in the large dimensions: $\Omega_0^2 \in \left\{0.998004, 0.992063,0.982318,0.968992\right\}$. Hence, although the dynamics of the string in the internal space is sensitively dependent on the number of windings, the apparent insensitivity of the $(3+1)$-dimensional dynamics to $m$ may be explained by the fact that $\Omega_0^2$ remains approximately constant for $m \in \left\{1,2,3,4\right\}$. From Eq. (\ref{7}), we see that the initial acceleration of the string in the radial direction of the compact space is determined by the condition $C^2 \lesseqgtr a^{-2}\rho_0^2R_0^4m^2[a^2(1-\dot{\rho}_0^2)-\dot{R}_0^2](a^2\rho_0^2 + m^2R_0^2)^{-2}$, which yields $\ddot{R}_0 \lesseqgtr 0$. With $\dot{\rho}_0 = \dot{R}_0 = 0$, this condition reduces to $C^2 \lesseqgtr \Omega_0^2(1-\Omega_0^2)/a^2$, and it is straightforward to check analytically that a transition from (initially) expanding windings to (initially) collapsing windings must occur between $m =2$ and $m =3$ in the examples considered. By contrast, in order to affect the macroscopic behavior of the string, we need to increase the value of $\Omega_0^2$ close to, or beyond, the critical value $\Omega_0^2 = 1/2$. In our examples, this corresponds to setting $m = 22.36$ but, since $m$ is an integer, a transition between initially expanding and initially contracting loops in $(3+1)$ dimensions must occur between $m = 22$ and $m = 23$. (It may be verified numerically that this is indeed the case.) 

\begin{figure}[h] 
\centering
\includegraphics[width=8cm]{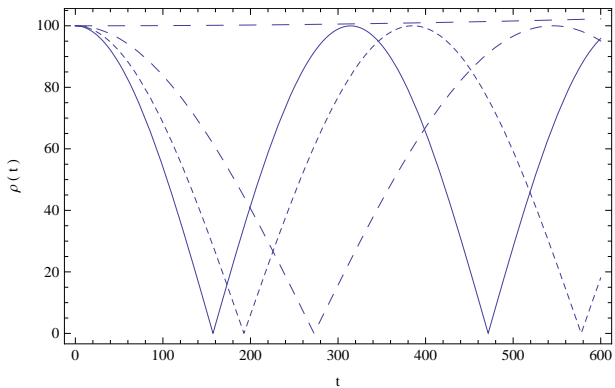}
\caption{Time variation of $\protect\rho(t)$ in Model I for $m=1 $, $a=1/\protect\sqrt{2}$, $C=0.05$, and for different values of $\dot{R}_0^2$:
$\dot{R}_0^2=0$ ({\it solid curve}), $\dot{R}_0^2=1/6$ ({\it dotted curve}), $\dot{R}_0^2=1/3$ ({\it short dashed curve}) and $\dot{R}_0^2=1/2$ ({\it long dashed curve}).} \label{f6}
\end{figure}

\begin{figure}[h] 
\centering
\includegraphics[width=8cm]{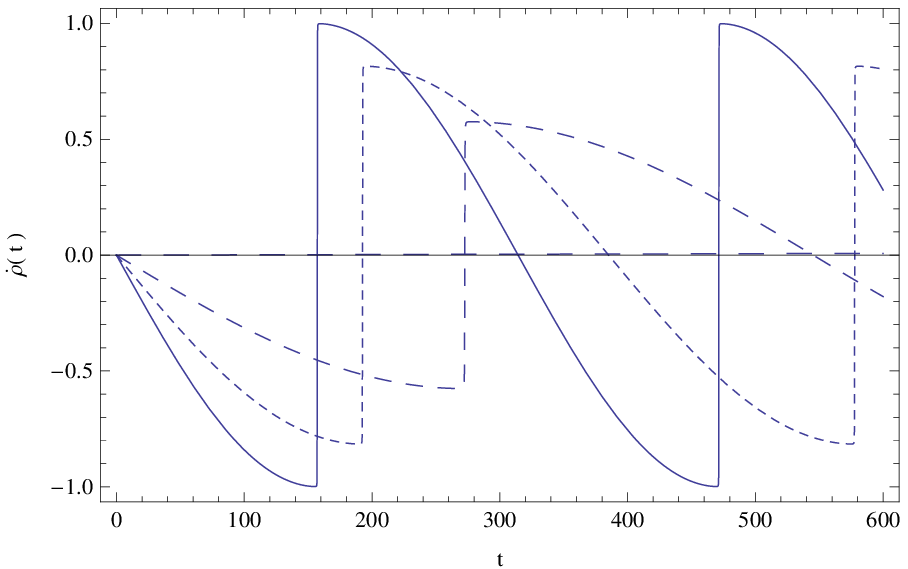}
\caption{Time variation of $\dot{\protect\rho}(t)$ in Model I for $m=1$, $a=1/\protect\sqrt{2}$, $C=0.05$, and for different values of $\dot{R}_0^2$:
$\dot{R}_0^2=0$ ({\it solid curve}), $\dot{R}_0^2=1/6$ ({\it dotted curve}), $\dot{R}_0^2=1/3$ ({\it short dashed curve}) and $\dot{R}_0^2=1/2$ ({\it long dashed curve}).} \label{f7}
\end{figure}

\begin{figure}[h] 
\centering
\includegraphics[width=8cm]{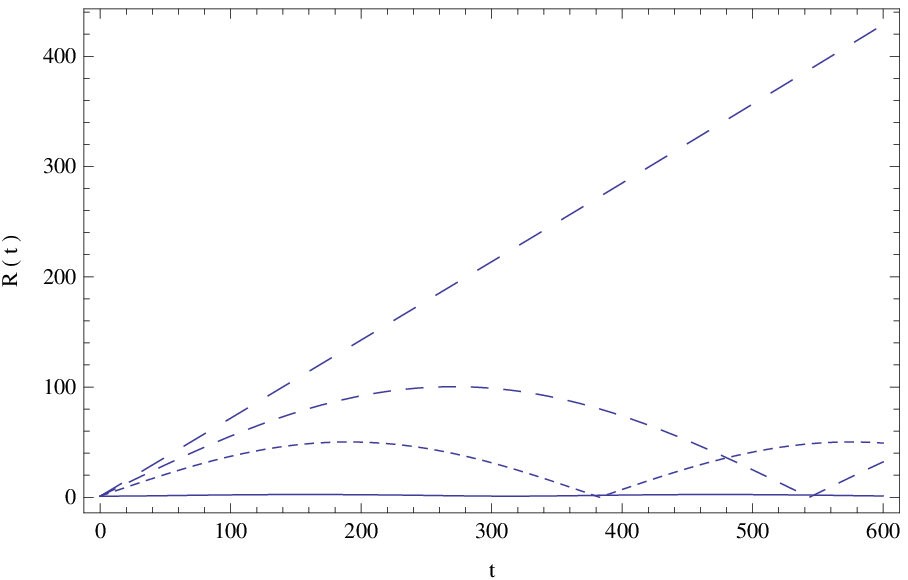}
\caption{Time variation of $R(t)$ in Model I for $m=1$, $a=1/\protect\sqrt{2}$, $C=0.05$, and for different values of $\dot{R}_0^2$:
$\dot{R}_0^2=0$ ({\it solid curve}), $\dot{R}_0^2=1/6$ ({\it dotted curve}), $\dot{R}_0^2=1/3$ ({\it short dashed curve}) and $\dot{R}_0^2=1/2$ ({\it long dashed curve}).} \label{f8}
\end{figure}

\begin{figure}[h] 
\centering
\includegraphics[width=8cm]{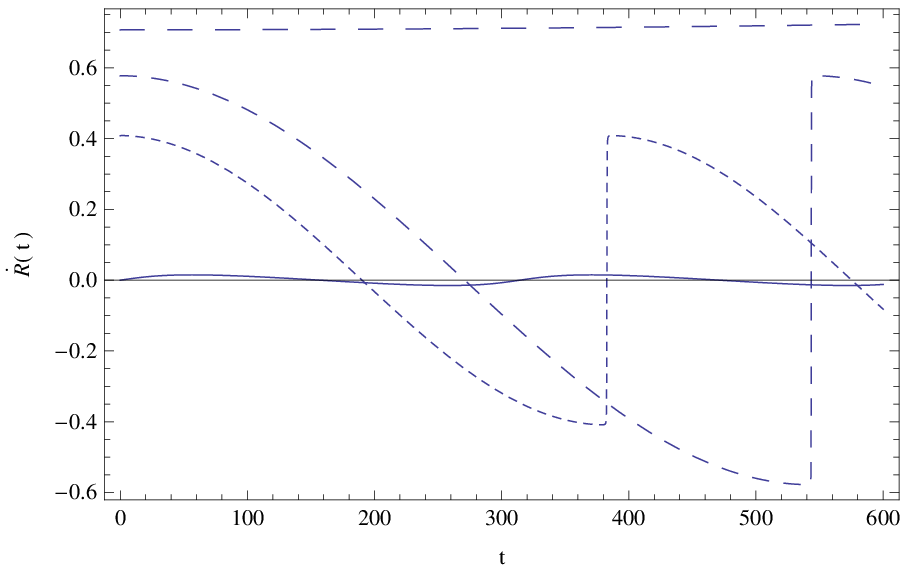}
\caption{Time variation of $\dot{R}(t)$ in Model I for $m=1$, $a=1/\protect\sqrt{2}$, $C=0.05$, and for different values of $\dot{R}_0^2$:
$\dot{R}_0^2=0$ ({\it solid curve}), $\dot{R}_0^2=1/6$ ({\it dotted curve}), $\dot{R}_0^2=1/3$ ({\it short dashed curve}) and $\dot{R}_0^2=1/2$ ({\it long dashed curve}).} \label{f9}
\end{figure}

\begin{figure}[h] 
\centering
\includegraphics[width=8cm]{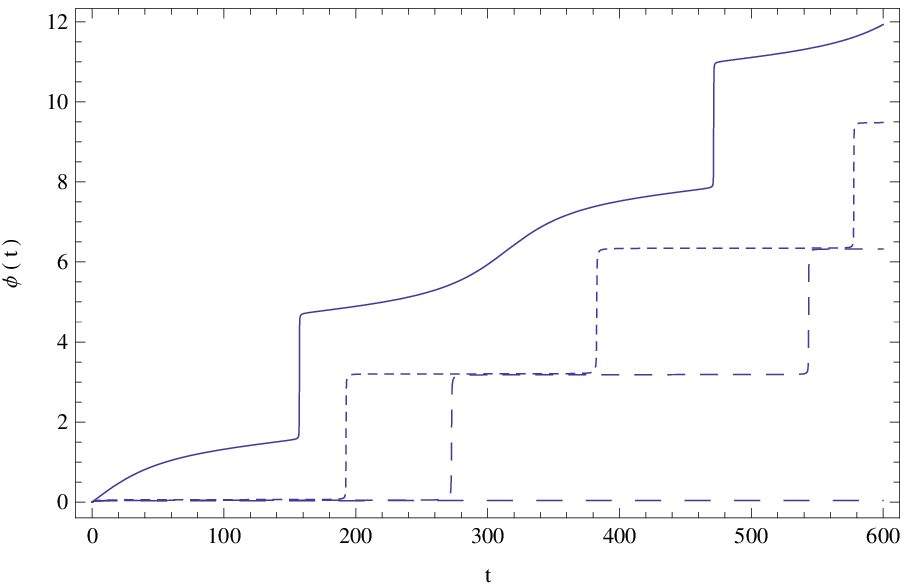}
\caption{Time variation of $\protect\phi(t)$ in Model I for $m=1$, $a=1/\protect\sqrt{2}$, $C=0.05$, and for different values of $\dot{R}_0^2$:
$\dot{R}_0^2=0$ ({\it solid curve}), $\dot{R}_0^2=1/6$ ({\it dotted curve}), $\dot{R}_0^2=1/3$ ({\it short dashed curve}) and $\dot{R}_0^2=1/2$ ({\it long dashed curve}).} \label{f10}
\end{figure}

However, it is more interesting to consider scenarios in which $\dot{R}_0^2 > 0$. For the sake of simplicity, we again assume $\dot{\rho}_0=0$, but consider the $m=1$ case with $\dot{R}_0^2 \in \left\{0,1/6,1/3,1/2\right\}$. The behavior of the functions $\rho$, $\dot{\rho}$, $R$, $\dot{R}$ and $\phi$, for these values, are plotted in Figs. \ref{f6}, \ref{f7}, \ref{f8}, \ref{f9} and \ref{f10}. For the sake of comparison, all other initial conditions are the same as in Figs. \ref{f1}, \ref{f2}, \ref{f3}, \ref{f4} and \ref{f5}.

The first and fourth examples, given by the solid and long dashed curves, respectively, correspond to the $R = const.$ ($\rho \neq const.$) and $\rho = const.$ ($R \neq const.$) solutions. The first of these may be described analytically by Eqs. (\ref{R=const_EOM1}) and (\ref{R=const_EOM2}) and corresponds to the analytic solution (\ref{R=const_EOM3}), found previously in \cite{Yamauchi:2014ita,LaYo12}, while the numerical solution corresponding to the final dashed curve implies that a ``mirror image" analytic solution also exists, in which $R$ and $\rho$ are interchanged. (This solution is derived explicitly in Sect. \ref{S3.2}.) In the $\rho = const.$ case, corresponding to $\dot{R}_0^2 = 1/2$, $R$ grows linearly in time without limit and, even for $\dot{R}_0^2 = 1/3$, the amplitude of the oscillations in $R$ is comparable to the initial $(3+1)$-dimensional loop radius. However, we must remember that, in reality, the extra dimensions must be compact, so that $R(t)$ obeys periodic boundary conditions which should be applied at some critical scale $\mathcal{R}$, the compatification radius of the internal space. Nonetheless, it is clear from these results that, were the string to be compactified on an flat internal space with trivial topology, the $(3+1)$-dimensional dynamics would be unaffected by oscillations in the internal space, regardless of their ``amplitude". For example, suppose we were to define our internal space simply by identifying opposing points on a flat two-dimensional disk, $D^2$, of radius $\mathcal{R}$, so that, having extended from $R = 0$ to $R = \mathcal{R}$, the winding radius simply retraced its previous path, from $R = \mathcal{R}$ to $R = 0$, and so on, \emph{ad infinitum}. In this case, we could replot the continuous dashed line in Fig. \ref{f8} as a saw-tooth dashed line (with a new cycle beginning every time $R(t) = \mathcal{R}$), but it would in no way affect the dynamics of $\rho(t)$ in the macroscopic dimensions. 

The important thing to note is that, if the amplitude of the string oscillations in the compact space becomes larger than the compactification radius, then the curvature and topology of the internal space \emph{cannot} be neglected. Though the amplitude of $R(t)$ in the examples considered in Figs. \ref{f1}, \ref{f2}, \ref{f3}, \ref{f4} and \ref{f5} remains small compared to the $(3+1)$-dimensional loop radius, those in Figs. \ref{f6}, \ref{f7}, \ref{f8}, \ref{f9} and \ref{f10} do not. In this case, it is also no longer reasonable to assume that they remain small compared to the compactification radius since, in general, $\rho \gg R$. Hence, nontrivial effects due to the topology and curvature of the compact space, which affect the higher-dimensional dynamics, may, {\it in principle}, also affect the dynamics of the string in the large visible dimensions. This motivates our work on $S^2$ compactification in Sect. \ref{S3.3} but, before then, we first determine the critical points of the first system of string EOM, Eqs. (\ref{6}) and (\ref{7}).

\subsection{Critical points of Model I} \label{S3.2}

The critical points of the system are defined by the
conditions $\ddot{\rho}=0$, $\dot{\rho}=0$, $\ddot{R}=0$, $\dot{R}=0$. Substituting these into Eqs. (\ref{6}) and (\ref{7}), we obtain the algebraic equations
\begin{equation}
\frac{a^{2}\rho ^{2}+m^{2}R^{2}}{\rho ^{2}R}=\pm \frac{a}{C},\ \ \ \frac{a^{2}\rho^{2}+m^{2}R^{2}}{\rho R^{2}}=\pm \frac{m}{C},  \label{8}
\end{equation}
and combining these expressions gives $C = \pm (1/2)R/a$, $mR = \pm a\rho$, which is clearly equivalent to $\Omega^2 = 1/2$. In terms of the constant $C$, the critical values of $m$ and $R$ may then be written explicitly as
\begin{equation}  \label{}
\rho = \pm 2mC, \ \ \ R = \pm 2aC.
\end{equation}

However, in general we may obtain a ``critical point", from a $(3+1)$-dimensional perspective, simply by setting $\rho=const$. The system then reduces to
\begin{equation} \label{R_EOM}
\dot{R}^2 - a^2 +C^2\left(\frac{a^2}{R} + \frac{m^2R}{\rho^2}\right)^2 = 0.
\end{equation}
Performing the substitution $y=R^2$, Eq. (\ref{R_EOM}) can be rewritten as
\begin{equation} \label{Euler1}
{\rm d}t = \pm \frac{1}{2}\frac{{\rm d}y}{\sqrt{-\alpha y^2 + \beta y -\gamma}},
\end{equation}
where
\begin{equation} \label{constants1}
\alpha = \frac{C^2m^4}{\rho^4}, \ \ \ Ê\beta = a^2\left(1 - \frac{2C^2m^2}{\rho^2}\right), \ \ \ Ê\gamma = C^2a^4.
\end{equation}
This may be integrated using an Eulerian substitution of the second kind \cite{LaWa10}. The method proceeds as follows. Firstly, let us assume that the quadratic equation $-\alpha y^2 + \beta y - \gamma = 0$ has two real roots, $A$ and $B$. We then define a dummy variable $u$ such that
\begin{equation}
\sqrt{-\alpha y^2 + \beta y - \gamma} = (y-A)u.
\end{equation}
Since $-\alpha y^2 + \beta y - \gamma = (y-A)(B-y)$, we then have
\begin{equation}
y = \frac{Au^2 + B\alpha}{u^2 + \alpha}, \ \ \ \frac{u}{\sqrt{\alpha}} = \pm \sqrt{\frac{B-y}{y-A}}.
\end{equation}
Combining these results gives
\begin{equation}
\int {\rm d}t = \pm \int \frac{{\rm d}u}{u^2 + \alpha} = \pm\frac{1}{\sqrt{\alpha}} \tan^{-1}\left(\frac{u}{\sqrt{\alpha}}\right) + K,
\end{equation}
where $K$ is an arbitrary integration constant, so that
\begin{equation}
y = B\left[1 + \left(\frac{A-B}{B}\right)\sin^2(\sqrt{\alpha}t + K)\right],
\end{equation}
and $R(t)$ takes the same functional form, for $\rho = const.$, that $\rho(t)$ takes when $R = const.$, as expected by symmetry. In terms of the model parameters, the roots of the quadratic equation are
\begin{equation}
y_{\pm} = \frac{1}{2}\frac{a^2\rho^2}{C^2m^4}\left(1 - \frac{2C^2m^2}{\rho^2}\right)\left[1 \pm \sqrt{1 - \frac{4C^4m^4}{\rho^4}\left(1 - \frac{2C^2m^2}{\rho^2}\right)^{-2}}\right]
\end{equation}
and their reality requires
\begin{equation} \label{puzzle-1}
C^2 \geq \frac{1}{4}\frac{\rho^2}{m^2} = \frac{1}{4}\frac{R^2}{a^2}\frac{\Omega}{\sqrt{1-\Omega^2}}.
\end{equation}
As the sum of the terms inside the square brackets is automatically positive, the reality of $R(t)$ then requires
\begin{equation}
C^2 \leq \frac{1}{2}\frac{\rho^2}{m^2} = \frac{1}{2}\frac{R^2}{a^2}\frac{\Omega}{\sqrt{1-\Omega^2}}.
\end{equation}
Combining these, we have
\begin{equation}
\frac{1}{4}\frac{R^2}{a^2}\frac{\Omega}{\sqrt{1-\Omega^2}} \leq C^2 \leq \frac{1}{2}\frac{R^2}{a^2}\frac{\Omega}{\sqrt{1-\Omega^2}},
\end{equation}
and the $\rho = const.$, $R = const.$ solution, found in \cite{Yamauchi:2014ita,LaYo12}, corresponds to saturation of the lower bound.

\subsection{Model II: numerical solutions of the EOM for windings on $S^2$} \label{S3.3}

In our second model, we assume that the background geometry contains an $S^2$ internal space of constant radius $\mathcal{R}$. Using cylindrical polars for the Minkowski part and canonical coordinates for $2$-sphere, the line element is
\begin{equation}
{\rm d}s^2 = a^2({\rm d}t^2 - {\rm d}\rho^2 - \rho^2{\rm d}\sigma^2 - {\rm d}z^2) - \mathcal{R}^2{\rm d}\theta^2 - \mathcal{R}^2\sin^2\theta {\rm d}\phi^2,
\end{equation}
where $\theta \in [0,\pi]$ is the polar angle and $\phi \in [0,2\pi)$ is the azimuthal angle. The ansatz for the circularly symmetric wound string loop is
\begin{equation}
X^{I} = (t=\xi \tau, \ \rho(t), \ \sigma, z=0, \ \theta(t), \ \phi(t,\sigma) = m\sigma + f(t))
\end{equation}
so that
\begin{equation}
(-\gamma) = \xi^2\left[(a^2(1-\dot{\rho}^2) - \mathcal{R}^2\dot{\theta}^2)(a^2\rho^2 + m^2\mathcal{R}^2\sin^2\theta) - a^2\rho^2\mathcal{R}^2\sin^2\theta\dot{\phi}^2\right].
\end{equation}
We assume $\rho_0 > 0$, $\theta_0 \neq 0,\pi$ since, in the limiting cases, the wound string ansatz becomes invalid. Setting the initial conditions such that $\rho_0 > 0$, $\theta_0 \neq 0,\pi$, the conservation of energy and momentum then implies $\rho(t) > 0$, $\theta(t) \neq 0, \pi$ for all time, which ensures that the EOM do not become singular.

The EOM of the second string model are:
\begin{equation}  \label{s21}
\ddot{\rho}+\frac{\rho}{a^{2}\rho^{2}+m^{2}\mathcal{R}^{2}\sin^{2}\theta}\left[a^{2}\left(1-\dot{\rho}^{2}\right) - \mathcal{R}^{2}\dot{\theta}^{2}-C^2\frac{(a^{2}\rho ^{2}+m^{2}\mathcal{R}^{2}\sin^{2}\theta)^2}{\rho^{4} \mathcal{R}^2\sin^{2}\theta}\right] =0,
\end{equation}

\begin{equation}  \label{s22}
\ddot{\theta} + \frac{\cos \theta \sin \theta}{a^{2}\rho^{2}+m^{2}\mathcal{R}^{2}\sin^{2}\theta} \left[m^2\left[a^{2}\left(1-\dot{\rho}^{2}\right) - \mathcal{R}^{2}\dot{\theta}^{2}\right]
-  a^2C^2\frac{(a^{2}\rho^{2}+m^{2}\mathcal{R}^{2}\sin^{2}\theta)^2}{\rho^{2} \mathcal{R}^4\sin^{4}\theta}\right] =0,
\end{equation}

\begin{equation}  \label{s23}
\dot{\phi} =  C \frac{a^{2}\rho^{2}+m^{2}\mathcal{R}^{2}\sin ^{2}\theta }{\rho^{2} \mathcal{R}^2\sin^{2}\theta },
\end{equation}
where $C$ again denotes an arbitrary integration constant, which admits the physical interpretation given in Eq. (\ref{C = l/E}). We note that, in this geometry, for small values of the polar angle $\theta(t) \ll 1$, the effective time-dependent radius of the windings is
\begin{equation}  \label{small_theta}
R(t) = \mathcal{R}\sin\theta(t) \approx \mathcal{R}\theta(t),
\end{equation}
and Eqs. (\ref{s21})-(\ref{s23}) reduce to Eqs. (\ref{3})-(\ref{5}) under this identification.

\begin{figure}[h] 
\centering
\includegraphics[width=8cm]{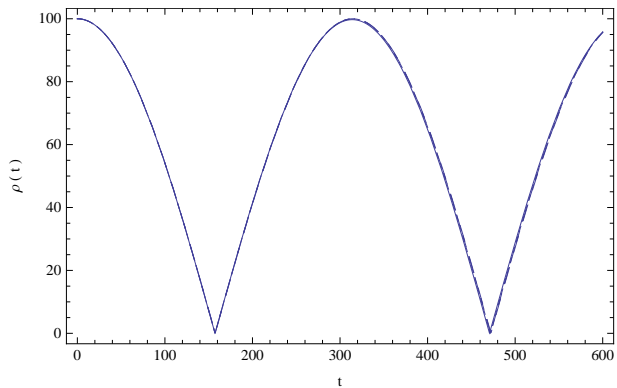}
\caption{Time variation of $\protect\rho(t)$ in Model II for different values of $m$: $m=1$ ({\it solid curve}), $m=2$ ({\it dotted curve}), $m=3$ ({\it short dashed curve}) and $m=4$
({\it long dashed curve}). As in Model I, the macroscopic dynamics of the string are remarkably insensitive to the number of higher-dimensional windings and the four curves are practically indistinguishable.}
\label{f11}
\end{figure}

\begin{figure}[h] 
\centering
\includegraphics[width=8cm]{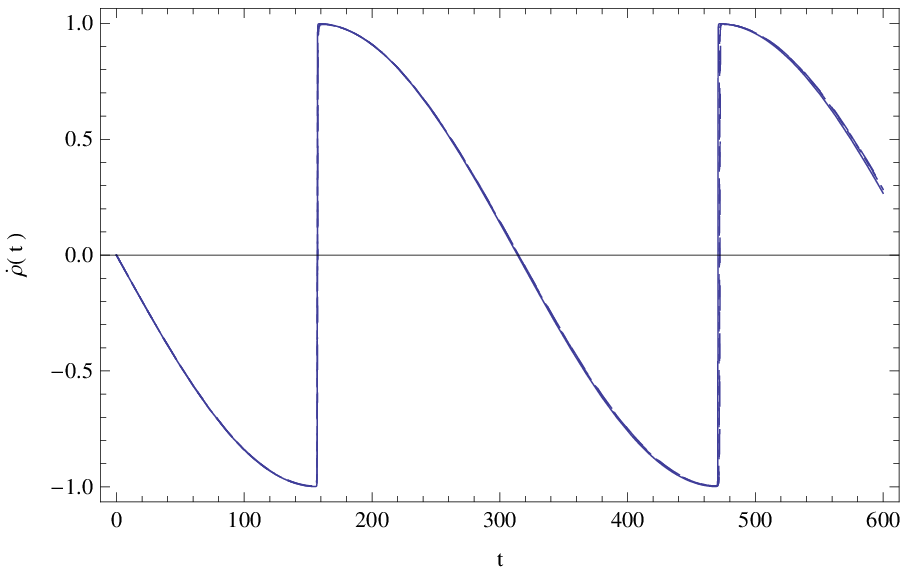}
\caption{Time variation of $\dot{\protect\rho}(t)$ in Model II for different values of $m$: $m=1$ ({\it solid curve}), $m=2$ ({\it dotted curve}), $m=3$ ({\it short dashed curve}) and $m=4$
({\it long dashed curve}). As in Fig. \ref{f11}, the four curves are practically indistinguishable due to the insensitivity of the macroscopic dynamics to the value of the winding number.}
\label{f12}
\end{figure}

\begin{figure}[h] 
\centering
\includegraphics[width=8cm]{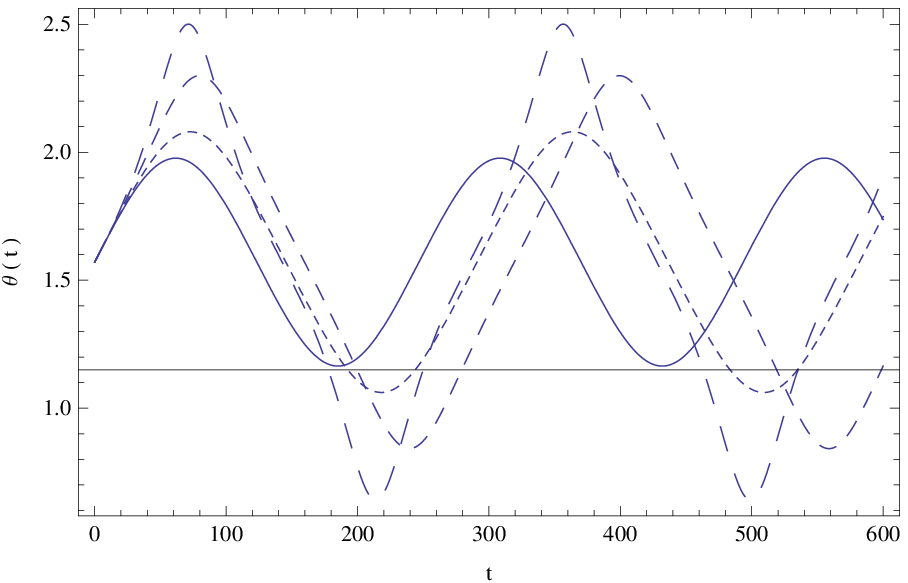}
\caption{Time variation of $\protect\theta(t)$ in Model II for different values of $m$: $m=1$ ({\it solid curve}), $m=2$ ({\it dotted curve}), $m=3$ ({\it short dashed curve}) and $m=4$
({\it long dashed curve}). This may be compared with Fig. \ref{f3}, which shows the behavior of $R(t)$ in Model I, though the two are only approximately equivalent in the limit $R(t) \equiv \mathcal{R}\sin\theta(t) \approx \mathcal{R}\theta(t)$, ($\theta(t) \ll 1$).}
\label{f13}
\end{figure}

\begin{figure}[h] 
\centering
\includegraphics[width=8cm]{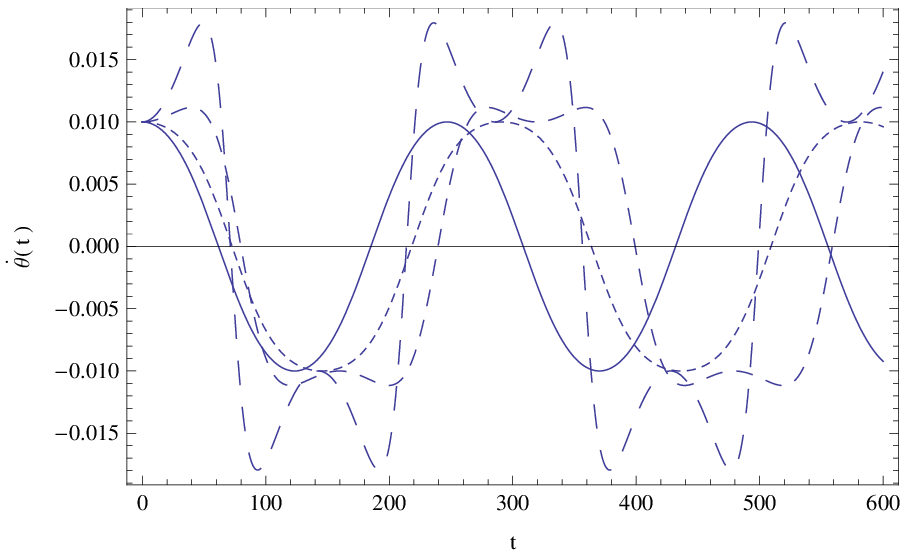}
\caption{Time variation of $\dot{\protect\theta}(t)$ in Model II for different values of $m$: $m=1$ ({\it solid curve}), $m=2$ ({\it dotted curve}), $m=3$ ({\it short dashed curve}) and $m=4$ ({\it long dashed curve}). This may be compared with Fig. \ref{f4}, which shows the behavior of $\dot{R}(t)$ in Model I, though the two are only approximately equivalent in the limit $R(t) \equiv \mathcal{R}\sin\theta(t) \approx \mathcal{R}\theta(t)$, ($\theta(t) \ll 1$).}
\label{f14}
\end{figure}

\begin{figure}[h] 
\centering
\includegraphics[width=8cm]{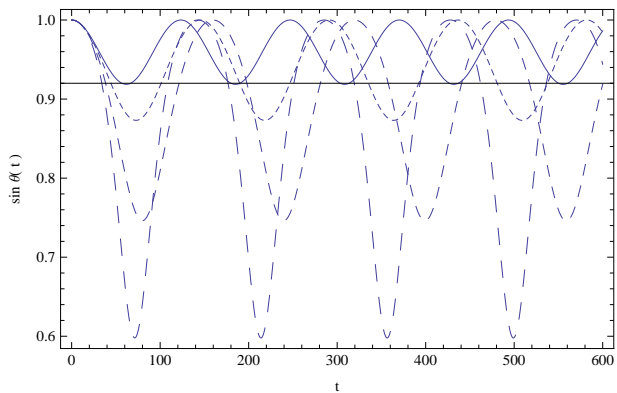}
\caption{Time variation of $\sin\protect\theta(t)$ in Model II for different values of $m$: $m=1$ ({\it solid curve}), $m=2$ ({\it dotted curve}), $m=3$ ({\it short dashed curve}) and $m=4$
({\it long dashed curve}). This is directly comparable with $R(t)$ in Model, shown in Fig. \ref{f3}, via the identification $R(t) \equiv \mathcal{R}\sin\theta(t)$.}
\label{f15}
\end{figure}

\begin{figure}[h] 
\centering
\includegraphics[width=8cm]{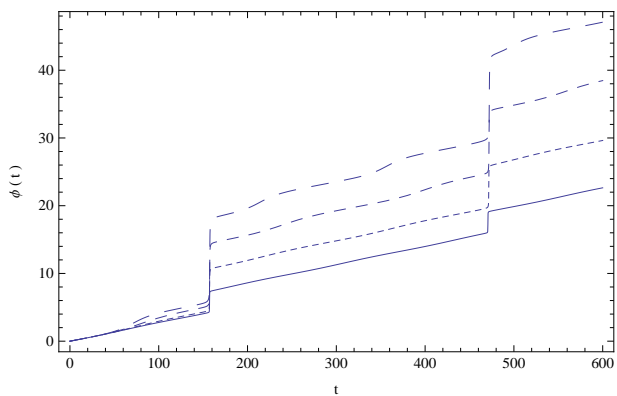}
\caption{Time variation of $\protect\phi(t)$ in Model II for different values of $m$: $m=1$ ({\it solid curve}), $m=2$ ({\it dotted curve}), $m=3$ ({\it short dashed curve}) and $m=4$
({\it long dashed curve}). As in Model I, we find that the period of $\protect\phi(t)$ matches that of the macroscopic loop radius $\rho(t)$, regardless of the detailed behavior of the effective winding radius $R(t) \equiv \mathcal{R}\sin\theta(t)$.}
\label{f16}
\end{figure}

Numerical solutions of the system of Eqs.~(\ref{s21})-(\ref{s23}) are presented in Figs. \ref{f11}, \ref{f12}, \ref{f13}, \ref{f14}, \ref{f15} and \ref{f16}, for $m \in \left\{1,2,3,4\right\}$, with $a = 1/\sqrt{2}$ and $C = 0.05$, as before. Specifically, $\rho$, $\dot{\rho}$, $\theta$, $\dot{\theta}$, $\sin\theta$ and $\phi$ are plotted in Figs. \ref{f11}, \ref{f12}, \ref{f13}, \ref{f14}, \ref{f15} and \ref{f16}, respectively. The initial conditions used are $\rho_0 =100$, $\dot{\rho}_0=0$ and $\phi_0=0$, as in Figs. \ref{f1}, \ref{f2}, \ref{f3}, \ref{f4}, \ref{f5}, \ref{f6}, \ref{f7}, \ref{f8}, \ref{f9} and \ref{f10}, together with $\mathcal{R}=1$, $\theta_0 =\pi/2$ and $\dot{\theta}^2_0=10^{-4}$. Thus, these examples are similar to those presented in Figs. \ref{f1}, \ref{f2}, \ref{f3}, \ref{f4} and \ref{f5} (in that they contain plots for multiple values of $m$), but also to those presented in Figs. \ref{f6}, \ref{f7}, \ref{f8}, \ref{f9} and \ref{f10} (in which $R_0=1$ and $\dot{R}_0^2 > 0$). However, in this case, the initial velocity of the string in the compact directions is small (and fixed to a single value).

From these plots, we see clearly that, although the microscopic behavior of $\mathcal{R}\theta(t)$, or even $\mathcal{R}\sin\theta(t)$, differs greatly from that of $R(t)$ in Model I, due to the nontrivial geometry and topology of the $S^2$, the behavior of $\phi(t)$ and of the macroscopic loop radius $\rho(t)$ remain stubbornly unresponsive to this change. Though the plots presented here represent only a small range of the possible parameter values and initial conditions, it may be verified that this is a general feature of the $(3+1)$-dimensional string dynamics.

\begin{figure}[h] 
\centering
\includegraphics[width=8cm]{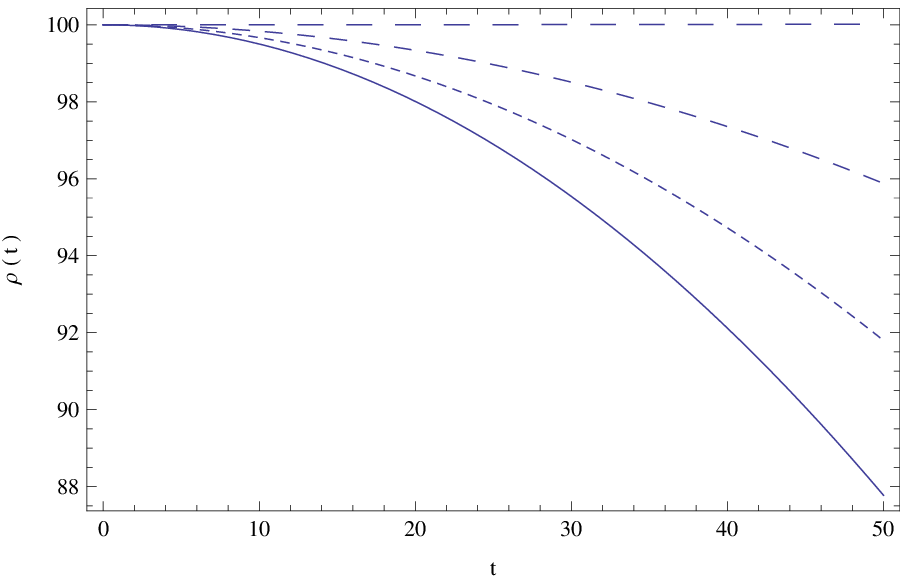}
\caption{Time variation of $\protect\rho(t)$ in Model II for $m = 1$ and different values of $\dot{\theta}_0^2$: $\dot{\theta}_0^2 = 0$ ({\it solid curve}), $\dot{\theta}_0^2 = 1/6$ ({\it dotted curve}), $\dot{\theta}_0^2 = 1/3$ ({\it short dashed curve}) and $\dot{\theta}_0^2 = 1/2$ ({\it long dashed curve}).}
\label{f11-A}
\end{figure}

\begin{figure}[h] 
\centering
\includegraphics[width=8cm]{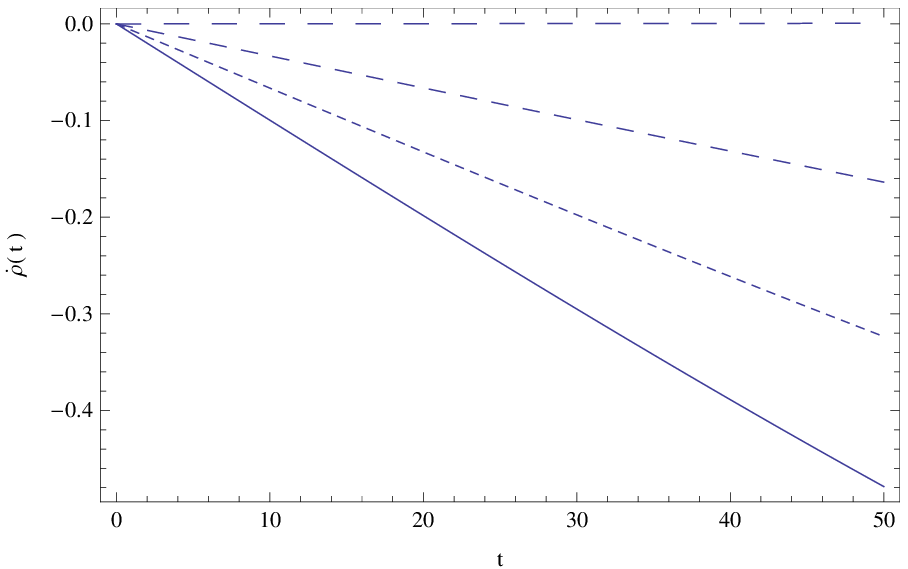}
\caption{Time variation of $\dot{\protect\rho}(t)$ in Model II for $m = 1$ and different values of $\dot{\theta}_0^2$: $\dot{\theta}_0^2 = 0$ ({\it solid curve}), $\dot{\theta}_0^2 = 1/6$ ({\it dotted curve}), $\dot{\theta}_0^2 = 1/3$ ({\it short dashed curve}) and $\dot{\theta}_0^2 = 1/2$ ({\it long dashed curve}).}
\label{f12-A}
\end{figure}

\begin{figure}[h] 
\centering
\includegraphics[width=8cm]{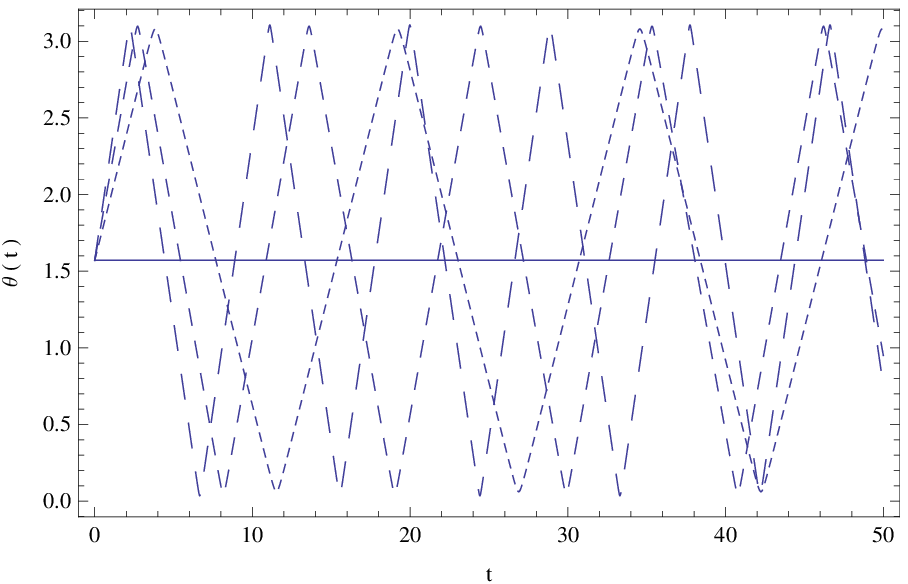}
\caption{Time variation of $\protect\theta(t)$ in Model II for $m = 1$ and different values of $\dot{\theta}_0^2$: $\dot{\theta}_0^2 = 0$ ({\it solid curve}), $\dot{\theta}_0^2 = 1/6$ ({\it dotted curve}), $\dot{\theta}_0^2 = 1/3$ ({\it short dashed curve}) and $\dot{\theta}_0^2 = 1/2$ ({\it long dashed curve}).}
\label{f13-A}
\end{figure}

\begin{figure}[h] 
\centering
\includegraphics[width=8cm]{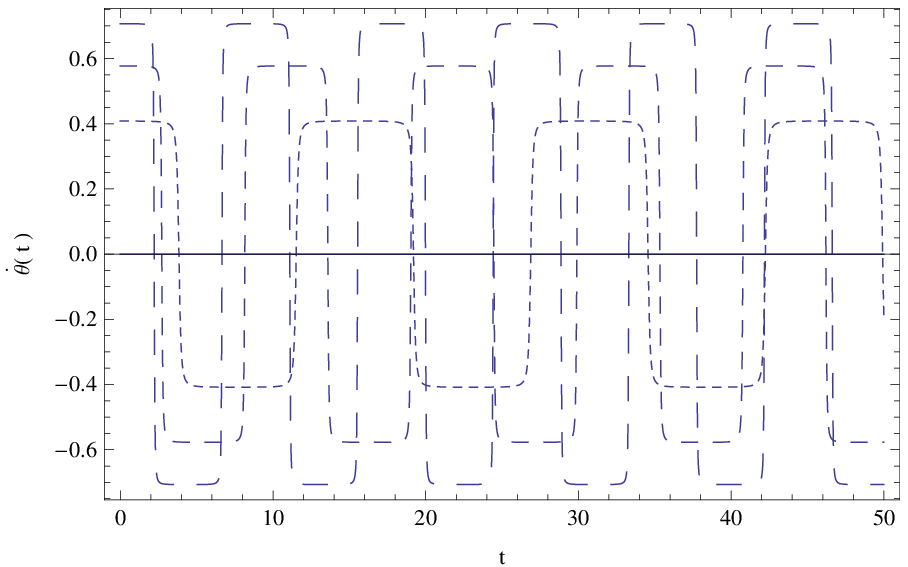}
\caption{Time variation of $\dot{\protect\theta}(t)$ in Model II for $m = 1$ and different values of $\dot{\theta}_0^2$: $\dot{\theta}_0^2 = 0$ ({\it solid curve}), $\dot{\theta}_0^2 = 1/6$ ({\it dotted curve}), $\dot{\theta}_0^2 = 1/3$ ({\it short dashed curve}) and $\dot{\theta}_0^2 = 1/2$ ({\it long dashed curve}).}
\label{f14-A}
\end{figure}

\begin{figure}[h] 
\centering
\includegraphics[width=8cm]{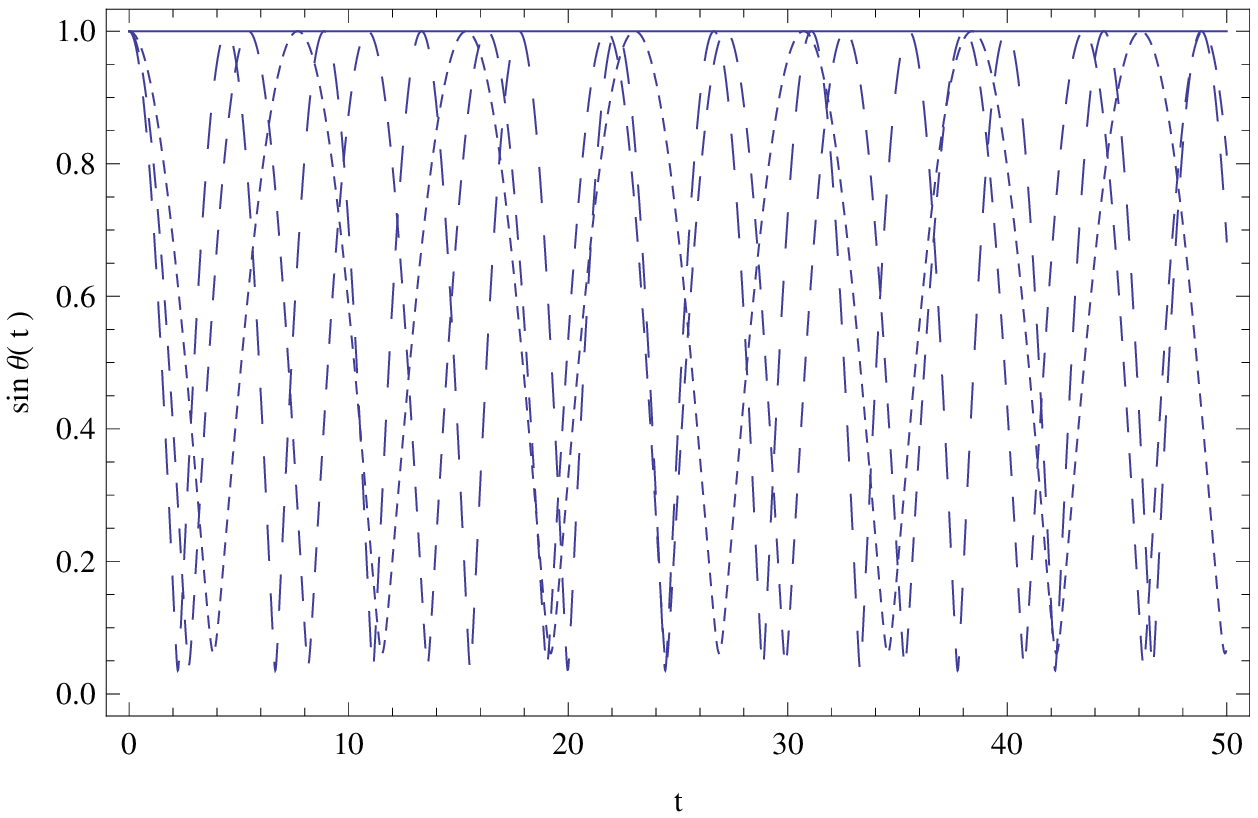}
\caption{Time variation of $\protect\sin\theta(t)$ in Model II for $m = 1$ and different values of $\dot{\theta}_0^2$: $\dot{\theta}_0^2 = 0$ ({\it solid curve}), $\dot{\theta}_0^2 = 1/6$ ({\it dotted curve}), $\dot{\theta}_0^2 = 1/3$ ({\it short dashed curve}) and $\dot{\theta}_0^2 = 1/2$ ({\it long dashed curve}).}
\label{f15-A}
\end{figure}

\begin{figure}[h] 
\centering
\includegraphics[width=8cm]{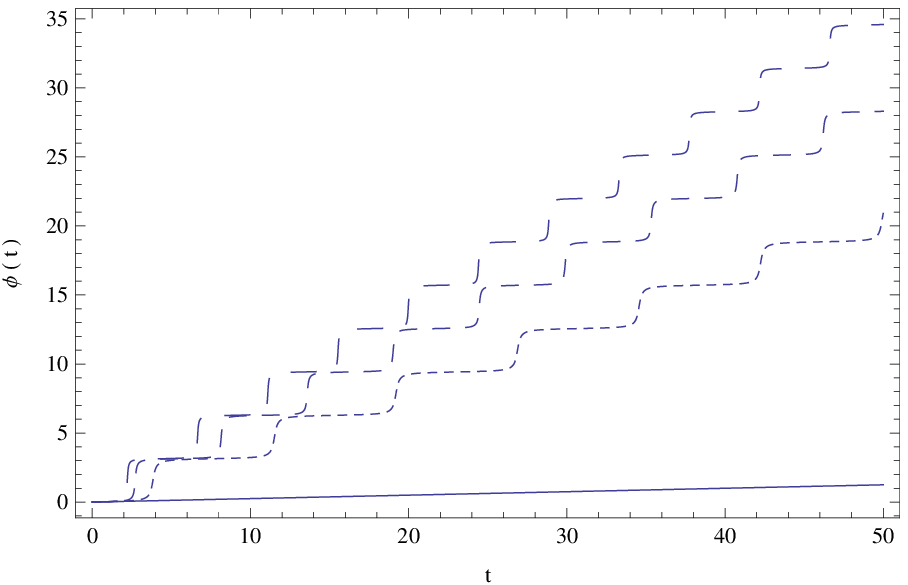}
\caption{Time variation of $\protect\phi(t)$ in Model II for $m = 1$ and different values of $\dot{\theta}_0^2$: $\dot{\theta}_0^2 = 0$ ({\it solid curve}), $\dot{\theta}_0^2 = 1/6$ ({\it dotted curve}), $\dot{\theta}_0^2 = 1/3$ ({\it short dashed curve}) and $\dot{\theta}_0^2 = 1/2$ ({\it long dashed curve}).}
\label{f16-A}
\end{figure}

As a further example of periodic behavior of the string in the compact space, that is qualitatively different, both from that of $R(t)$ in Model I and that shown in Figs. \ref{f11}, \ref{f12}, \ref{f13}, \ref{f14}, \ref{f15} and \ref{f16}, $\rho$, $\dot{\rho}$, $\theta$, $\dot{\theta}$, $\sin\theta$ and $\phi$ are plotted for $a = 1/\sqrt{2}$, $C=0.05$, $m=1$, $\rho_0=100$, $\dot{\rho}_0=0$, $\theta_0 = \pi/2$, $\phi_0=0$ and with $\dot{\theta}_0^2 \in \left\{0,1/6,1/3,1/2\right\}$ in Figs. \ref{f11-A}, \ref{f12-A}, \ref{f13-A}, \ref{f14-A}, \ref{f15-A} and \ref{f16-A}. For the sake of clarity, these plots cover a much shorter time range, which is less than one oscillation period in $\rho(t)$. However, yet again, it can be seen that the motion of the string in the large dimensions is extremely insensitive to detailed dynamics of the string in the internal space, so that the periodicity in the effective winding radius does not induce additional fluctuations in the expansion and contraction of the $(3+1)$-dimensional loop radius.

\begin{figure}[h] 
\centering
\includegraphics[width=8cm]{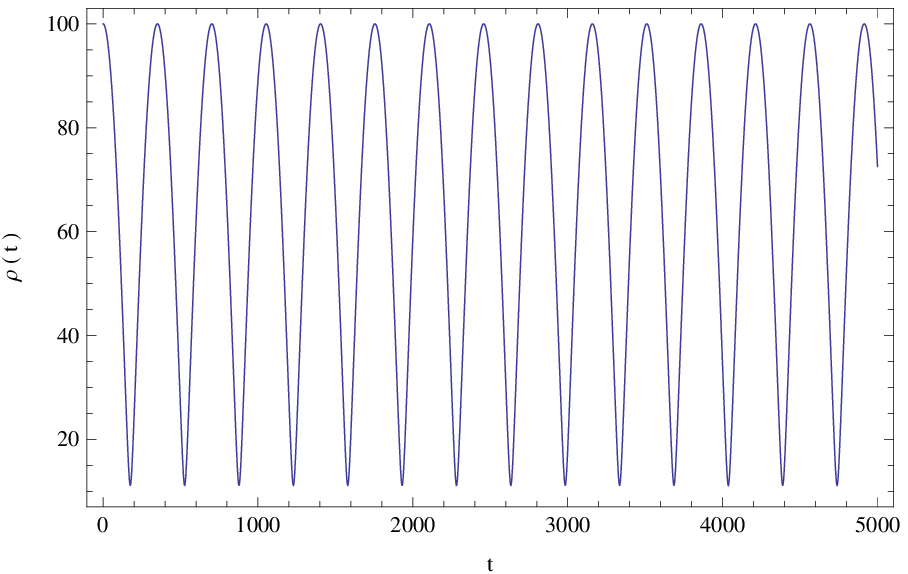}
\caption{Late time variation of $\rho(t)$ in Model II for $C = 0.5$ and $m = 20$.}
\label{f17}
\end{figure}

\begin{figure}[h] 
\centering
\includegraphics[width=8cm]{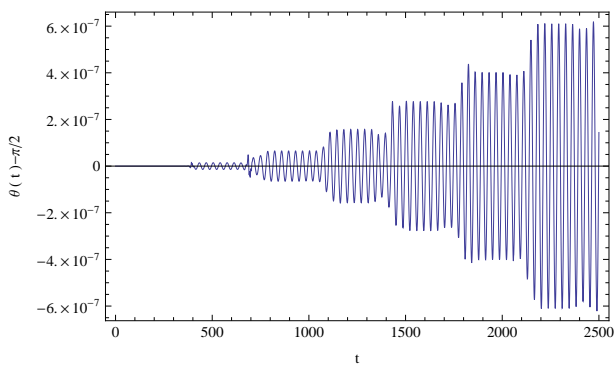}
\caption{Late time variation of $\protect\theta(t)$ in Model II for $C = 0.5$ and $m = 20$. The amplitude of oscillations around the initial value $\theta_0 = \pi/2$, which corresponds to windings around great circles in the $S^2$, grows {\it almost} monotonically with time, except for small scale fluctuations over (roughly) each half-oscillation in the loop radius $\rho(t)$, but remain small over long time periods.}
\label{f18}
\end{figure}

\begin{figure}[h] 
\centering
\includegraphics[width=8cm]{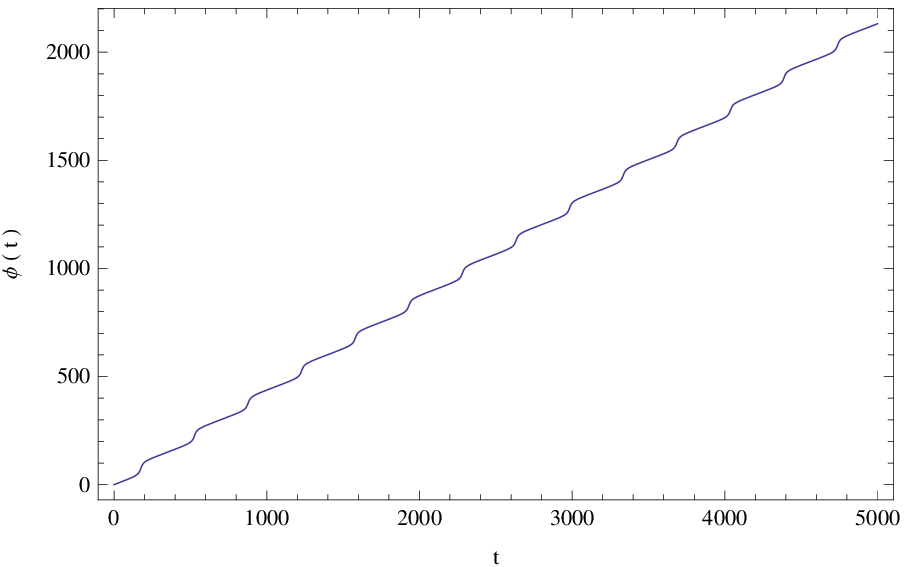}
\caption{Late time variation of $\protect\phi(t)$ in Model II for $C = 0.5$ and $m = 20$.}
\label{f19}
\end{figure}

A possible exception to this rule may appear only at very late times: Figs. \ref{f17}, \ref{f18} and \ref{f19}, show $\rho$, $\theta$ and $\phi$, obtained using the initial conditions $\rho_0 =100$, $\dot{\rho}_0=0$, $\theta_0 = \pi/2$, $\dot{\theta}_0 = 0$ and $\phi_0=0$, for $a = 1/\sqrt{2}$, but with $C = 0.5$ and $m = 20$. The time range considered is approximately one order of magnitude larger than that shown in Figs. \ref{f1}, \ref{f2}, \ref{f3}, \ref{f4}, \ref{f5}, \ref{f6}, \ref{f7}, \ref{f8}, \ref{f9}, \ref{f10}, \ref{f11}, \ref{f12}, \ref{f13}, \ref{f14}, \ref{f15} and \ref{f16}. From these figures, we see that the initial condition $\theta_0=\pi/2$ holds for a considerable length of time but, due to the strongly nonlinear nature of the EOM, oscillations in $\theta(t)$ begin after a finite time interval and, thereafter, their amplitude increases without limit. (The model parameters in this example are chosen so as to illustrate this behavior, but it may again be verified that it is characteristic of the system for a wide range of initial conditions and different values of the string constants of motion.) However, because of their extremely small initial magnitude (in our units, $\ll 10^{-6}$), extremely long time periods are required for any significant effect to occur in the large dimensions. Thus, oscillations in $\rho(t)$ appear completely regular over many cycles.

Due to computational limitations, it has not been possible to solve the string EOM up to time scales of order $10^4$ or above, in our chosen units. But, assuming that, for this set of parameters, the oscillations in $\theta(t)$ grow by at least one order of magnitude over this time period (as Fig. \ref{f18} suggests), a time interval of order $10^{24}$ would be required for them to reach an amplitude of approximately one radian. Thereafter, the $(3+1)$-dimensional dynamics of the string may indeed be significantly affected by its motion in the higher-dimensional space but, at late times, it is likely (physically) that a significant fraction of the original string mass would have been lost via gravitational wave emission, so that the assumption of conservation of energy and momentum used to obtain the string EOM (\ref{s21})-(\ref{s23}) becomes invalid. We may therefore conclude that, even if they are technically present, any imprints of the higher-dimensional string dynamics appearing in the effective $(3+1)$-dimensional motion are, in all probability, subdominant to effects produced by the interaction of the string with its local $(3+1)$-dimensional environment.

\subsection{Critical points of Model II} \label{S3.4}

The critical points of Eqs. (\ref{s21}) and (\ref{s22}) are defined by the conditions $\ddot{\rho}=0$, $\dot{\rho}=0$ and $\ddot{\theta}=0$, $\dot{\theta}=0$. For $\theta \neq \pi/2$, these can be obtained as solutions of the algebraic system
\begin{equation}  \label{cp21}
\frac{a^{2}\rho^{2}+m^{2}\mathcal{R}^{2}\sin^{2}\theta}{\rho^{2} \mathcal{R}\sin\theta} = \pm \frac{a}{C}, \ \ \ \frac{a^{2}\rho^{2}+m^{2}\mathcal{R}^{2}\sin^{2}\theta}{\rho \mathcal{R}^2\sin^{2}\theta} = \pm \frac{m}{C}.
\end{equation}
Following \cite{Yamauchi:2014ita}, we now define the local fraction of the string in the large dimensions $\omega^2(t,\sigma)$ as
\begin{equation}  \label{}
\omega^2 = \frac{a^{2}\rho^{2}}{a^{2}\rho^{2}+m^{2}R_{0}^{2}\sin^{2}\theta},
\end{equation}
so that the parameter $\Omega^2(t)$ is defined via
\begin{equation}  \label{}
\Omega^{-2} = \frac{1}{2\pi}\int_{0}^{2\pi}\omega^{-2}{\rm d}\sigma = \frac{1}{2\pi}\int_{0}^{2\pi}\frac{a^{2}\rho^{2}+m^{2}R_{0}^{2}\sin^{2}\theta}{a^2\rho^{2}}{\rm d}\sigma.
\end{equation}
The critical points of the system are given by $C = \pm(1/2)\mathcal{R}\sin\theta/a$, $m\mathcal{R}\sin\theta = \pm a\rho$, which corresponds to $\omega^2 = 1/2$. In \cite{Yamauchi:2014ita}, it was proved that wound strings with constant winding radius ($R = const.$), but arbitrary configurations in the large dimensions, are effectively tensionless, from a $(3+1)$-dimensional perspective, when $\omega^2(t,\sigma) = 1/2$ for all $\sigma$ and $t$. The result above is therefore interesting, as it implies that this condition may be even more general, extending to at least some models for which the effective winding radius is a function of time (for example, $R(t) \equiv \mathcal{R}\sin\theta(t)$, as in Eq. (\ref{small_theta}), for this model), for strings wrapping $S^1$ subcycles in more complex simply connected internal manifolds.

In terms of the constant $C$, the critical values of $\rho$ and $\theta$ may be written explicitly as
\begin{equation}  \label{crit-1}
\rho = \pm 2mC, \ \ \ \theta = \pm \sin^{-1}\left(\frac{2aC}{\mathcal{R}}\right), \ \ (C^2 \neq (1/4)\mathcal{R}^2/a^2).
\end{equation}
However, the limiting case for which $C \rightarrow (1/2)\mathcal{R}/a$, $m\mathcal{R} \rightarrow \pm a\rho$ and $\theta \rightarrow \pi/2$, which corresponds to the maximum effective winding radius, with the string wrapping great circles in the $S^2$, is explicitly \emph{not} included in this analysis.

For $\theta = \pi/2$, $\theta$-dependence vanishes completely from Eqs. (\ref{s21}) and (\ref{s22}). In this case, Eq. (\ref{s22}) becomes trivial (being merely an identity) and Eq. (\ref{s21}) yields
\begin{equation}  \label{}
\frac{a^{2}\rho^{2}+m^{2}\mathcal{R}^{2}}{\rho^{2} \mathcal{R}} = \pm \frac{a}{C} \iff \rho^2 = \left(\frac{C}{\pm \mathcal{R} - aC}\right)\frac{m^2\mathcal{R}^2}{a}.
\end{equation}
The positivity of $\rho^2$ then requires either $C < \mathcal{R}/a$ (for the positive solution) or $C > -\mathcal{R}/a$ (for the negative solution). Equivalently, we may require $C^2 < \mathcal{R}^2/a^2$. Overall, there are four critical points, given by
\begin{equation}  \label{crit-2}
\rho = \pm \sqrt{\frac{C}{\pm \mathcal{R} - aC}}\frac{m\mathcal{R}}{\sqrt{a}}, \ \ \ \theta = \frac{\pi}{2},
\end{equation}
where we fix $a > 0$, on physical grounds, to ensure the reality of the loop radius. (As we shall see in Sect. \ref{S5.2}, this is also necessary to ensure the positivity of the Hamiltonian.) The positivity of $\rho$ then requires us to choose the positive sign in from of the square root in Eq. (\ref{crit-2}) when $m>0$ and the negative sign when $m<0$.

If we require consistency with Eq. (\ref{crit-1}) in the limit $\theta \rightarrow (\pi/2)^{\pm}$ (i.e., if we require that there is no discontinuity in the critical points of the system as $C^2 \rightarrow (1/4)\mathcal{R}^2/a^2$), then we may set $C = \pm (1/2)\mathcal{R}/a$ in Eq. (\ref{crit-2}), which then yields $m\mathcal{R} = \pm a\rho$. In this case, the second set of critical points may be written in an equivalent form as
\begin{eqnarray} \label{crit-3}
\rho = \pm 2mC, \ \ \ \theta = \frac{\pi}{2}, \ \ (C^2 = (1/4)\mathcal{R}^2/a^2),
\end{eqnarray}
but we will leave them in the more general form given by Eq. (\ref{crit-2}), when performing the stability analysis.

Again, we may also consider a more general ``critical point", from a $(3+1)$-dimensional perspective, by setting $\rho = const.$ for $\theta = \theta(t)$. The system then reduces to
\begin{equation} \label{theta_EOM}
\mathcal{R}^{2}\dot{\theta}^{2} - a^{2}  + C^2\left(\frac{a^{2}}{\mathcal{R}\sin\theta} + \frac{m^2\mathcal{R}\sin\theta}{\rho^2}\right)^2 = 0.
\end{equation}
This equation is separable and may be written in the form,
\begin{equation}\label{Euler2}
{\rm d}t = \pm \frac{1}{2}\frac{{\rm d}z}{\sqrt{\delta z^3 - \tilde{\alpha} z^2 + \tilde{\beta} z -\gamma}},
\end{equation}
where $z = \mathcal{R}^2\sin^2\theta$ and
\begin{equation} \label{constants2}
\delta = \frac{C^2m^4}{\mathcal{R}^2\rho^4}, \ \tilde{\alpha} = \left(\frac{a^2}{\mathcal{R}^2}-\frac{2C^2a^2m^2}{\mathcal{R}^2\rho^2 } + \frac{C^2m^4}{\rho^4}\right), \ \tilde{\beta} = \left(a^2 + \frac{C^2a^4}{\mathcal{R}^2} - \frac{2C^2a^2m^2}{\rho^2}\right), \ \gamma = C^2a^4,
\end{equation}
then integrated numerically. For
\begin{equation}  \label{}
m^2\mathcal{R}^2 \ll 2a^2\rho^2, \ \ \  C^2 \ll \frac{a^2}{\mathcal{R}^2}\frac{\rho^4}{m^4},
\end{equation}
we have $\tilde{\alpha} \rightarrow \alpha$, $\tilde{\beta} \rightarrow \beta$, where $\alpha$ and $\beta$ are defined in Eq. (\ref{constants1}). If, in addition, we have $\sin\theta \ll a^2\rho^2/(m^2\mathcal{R}^2)$, which ensures that $\sin\theta \ll 1$ when $\Omega^2 \big|_{\theta ={\pi/2}} \equiv a^2\rho^2/(a^2\rho^2 + m^2\mathcal{R}^2) \leq 1/2$, the $\delta z^3$ term can be neglected and $z \approx y \approx \mathcal{R}^2\theta^2$. In this case, we recover the previous expression Eq. (\ref{Euler1}) as an approximate EOM for the system and an analytic solution can be obtained.

\section{Kosambi-Cartan-Chern (KCC) theory and Jacobi stability} \label{S4}

In the present Section we very briefly summarize the basic concepts and results of the KCC theory (for a detailed presentation see \cite{An00} and \cite{rev}).

Let $\mathcal{M}$ be a real, smooth, $n$-dimensional manifold, and let $(T\mathcal{M}, \pi, \mathcal{M})$ be its tangent bundle, where $\pi : T\mathcal{M} \rightarrow \mathcal{M}$ is a projection from the total space $T\mathcal{M}$ to the base manifold $\mathcal{M}$. For a local chart $\left(U, \phi = \left(x^i\right)\right)$ on $\mathcal{M}$, where ${\bf x}=\left( x^{i}\right) =\left(x^{1},x^{2}, \dots ,x^{n}\right) $, we denote by $\left(\pi ^{-1}(U),\Phi = \left(x^i, y^i=\dot{x}^i\right)\right)$ the induced local chart $T\mathcal{M}$, where $\dot{x}^i={\rm d}x^i/{\rm d}t$, ${\bf y}=\left( y^{i}\right) =\left(y^{1},y^{2}, \dots ,y^{n}\right) $, and $t$ is the time coordinate, introduced so that $\left(t, x^i\right)$ are defined on an open subset of the trivial bundle $\mathbb{R} \times \mathcal{M}$ with base manifold $\mathbb{R}$ and fibre $\mathcal{M}$. It is assumed  that the coordinate $t$ on $\mathbb{R}$ is kept fixed, while arbitrary coordinate transformations on the fibre $\mathcal{M}$ are allowed. Therefore, in the framework of the KCC theory, the only admissible coordinate transformations on $\mathcal{M}$ are
\begin{equation}
\tilde{t}=t, \ \ \ \tilde{x}^{i}=\tilde{x}^{i}\left( x^{1},  x^{2}, \dots ,x^{n}\right), \ Êi\in \left\{1 ,2, \dots ,n\right\} .  \label{ct}
\end{equation}

Vector fields $S$ of the form
\be
S=y^i\frac{\partial }{\partial x^i}-2G^i\left({\bf x},{\bf y},t\right)\frac{\partial }{\partial y^i},
\ee
defined on $T\mathcal{M}$, are called {\it semisprays} \cite{S0,S1,S2}. The functions ${\bf G}$, with components $G^i({\bf x}, {\bf y},t)$, defined on domains of local charts, are {\it the local coefficients} of the semispray.  The vector field $S$ is called a {\it spray} in the particular case in which the coefficients $G^i = G^i \left({\bf x}, {\bf y}\right)$ are {\it homogeneous of degree two in ${\bf y}$}.

We define a {\it path} or a {\it geodesic} of the semispray $S$ as a curve $c : t \rightarrow {\bf x}(t)$ on $\mathcal{M}$. The geodesic on a semispray  has the property that its lift $c' : t\rightarrow \left(x^i(t), \dot{x}^ i(t)\right)$ to $T\mathcal{M}$ is an integral curve of $S$, that is, it satisfies the equation \cite{S0,S1,S2}
\be\label{EM0}
\frac{{\rm d}^2{\bf x}}{{\rm d}t^2}+2{\bf G}\left({\bf x},{\bf y},t\right)=0.
\ee
Conversely, for any system of ordinary differential equations of the form (\ref{EM0}), which is globally defined, the functions ${\bf G}$ define a semispray on $T\mathcal{M}$ \cite{S1,S2}.

In many situations, the EOM of a physical system can be derived from a Lagrangian $L=L\left(x^i,y^i,t\right)$ via the Euler-Lagrange equations,
\begin{equation} \label{EL}
\frac{{\rm d}}{{\rm d}t}\frac{\partial L}{\partial y^{i}}-\frac{\partial L}{\partial
x^{i}}=f_{i}, \ \ i=1,2, \dots ,n,
\end{equation}
where $f_{i}$, $i \in \left\{1,2, \dots ,n\right\}$, are the components of the external force. We call the triplet $\left(M,L,f_{i}\right) $ {\it a Finslerian mechanical system} \cite{MHSS, MiFr05}. For a regular Lagrangian $L$, the Euler-Lagrange equations introduced in Eq.~(\ref{EL}) are equivalent to a system of second-order ordinary (usually strongly nonlinear) differential equations
\begin{equation} \label{EM}
\frac{{\rm d}^{2}x^{i}}{{\rm d}t^{2}}+2F^{i}\left( x^{j},y^{j},t\right) =0,
\end{equation}
where each function $F^{i}\left( x^{j},y^{j},t\right) $ is $C^{\infty }$ in a neighborhood of some initial conditions $\left(\left(x\right)_{0},\left( y\right) _{0},t_{0}\right)$.

The basic idea of KCC theory is to start from an arbitrary system of second-order differential equations of the form (\ref{EM0}), defined on the base space $\mathcal{M}$, with no \textit{a priori} Lagrangian function assumed, and {\it to study the behavior of its trajectories on $\mathcal{M}$ by analogy with the trajectories of the Euler-Lagrange system (\ref{EM})}.

To analyze the geometry associated with the dynamical system defined by Eq.~(\ref{EM0}), as a first step, we introduce a nonlinear connection ${\bf N}$ on $\mathcal{M}$, with coefficients $N_{j}^{i}$, defined as \cite{MHSS}
\begin{equation}
N_{j}^{i}\left(x^i,y^i,t\right)=\frac{\partial G^{i}\left(x^i,y^i,t\right)}{\partial y^{j}}.
\end{equation}
The nonlinear connection can be understood geometrically in terms of a dynamical covariant derivative $\nabla ^N$ as follows \cite{Punzi}: let two vector fields $v$, $w$ be defined over a manifold $M$. Then  the covariant derivative $\nabla ^N$ on $\mathcal{M}$ is defined as
\begin{equation}  \label{con}
\nabla _v^Nw=\left[v^j\frac{\partial }{\partial x^j}w^i+N^i_j(x,y)w^j\right]%
\frac{\partial }{\partial x^i}.
\end{equation}
For $N_i^j(x,y)=\Gamma _{il}^j(x)y^l$, Eq.~(\ref{con}) reduces to the definition of the covariant derivative for the special case of a standard linear connection, as defined in Riemmannian geometry.

For the non-singular coordinate transformations, introduced through Eq.~(\ref{ct}), we define the KCC-covariant differential of an arbitrary vector field $\theta^{i}\left(x^i,t\right)$ on the open subset $\chi \subseteq \mathbb{R}^{n}\times \mathbb{R}^{n}\times \mathbb{R}^{1}$ as \cite{An00,Sa05,Sa05a,An93}
\begin{equation}
\frac{D\theta ^{i}}{{\rm d}t}=\frac{{\rm d}\theta ^{i}}{{\rm d}t}+N_{j}^{i}\theta ^{j}.  \label{KCC}
\end{equation}
For $\theta ^{i}=y^{i}$, we obtain
\begin{equation}
\frac{Dy^{i}}{{\rm d}t}=N_{j}^{i}y^{j}-2G^{i}=-\epsilon ^{i}.
\end{equation}
The contravariant vector field $\epsilon^{i}$ on $\chi $ is called the first KCC invariant.

We now vary the trajectories $x^{i}(t)$ of the system (\ref{EM}) into nearby ones according to
\begin{equation}
\tilde{x}^{i}\left( t\right) =x^{i}(t)+\eta \xi ^{i}(t),  \label{var}
\end{equation}
where $\left| \eta \right| $ is a small parameter and $\xi ^{i}(t)$ are the components of a contravariant vector field defined along the path $x^{i}(t)$. Substituting Eq. (\ref{var}) into Eq. (\ref{EM}) and taking the limit $\eta \rightarrow 0$, we obtain the deviation equations in the form \cite{An00,Sa05,Sa05a,An93}
\begin{equation}
\frac{{\rm d}^{2}\xi ^{i}}{{\rm d}t^{2}}+2N_{j}^{i}\frac{{\rm d}\xi ^{j}}{dt}+2\frac{\partial G^{i}}{\partial x^{j}}\xi ^{j}=0.  \label{def}
\end{equation}
Equation (\ref{def}) can be reformulated in covariant form with the use of the KCC-covariant differential as
\begin{equation}
\frac{D^{2}\xi ^{i}}{{\rm d}t^{2}}=P_{j}^{i}\xi ^{j},  \label{JE}
\end{equation}
where we have denoted
\begin{equation}
P_{j}^{i}\left(t,x^i,y^i\right)=-2\frac{\partial G^{i}}{\partial x^{j}}-2G^{l}G_{jl}^{i}+ y^{l}\frac{\partial N_{j}^{i}}{\partial x^{l}}+N_{l}^{i}N_{j}^{l}+\frac{\partial N_{j}^{i}}{\partial t},
\end{equation}
and have introduced the Berwald connection $G_{jl}^{i}$, defined as \cite{An00,rev,Sa05,Sa05a,An93,MHSS}
\begin{equation}
G_{jl}^{i}\left(t,x^i,y^i\right)\equiv \frac{\partial N_{j}^{i}}{\partial y^{l}}.
\end{equation}
$P_{j}^{i}$ is called the second KCC-invariant, or the deviation curvature tensor, while Eq.~(\ref{JE}) is called the Jacobi equation. When the system (\ref{EM}) describes the geodesic equations, Eq.~(\ref{JE}) is the Jacobi
field equation, in either Riemann or Finsler geometry. The trace $P$ of the curvature deviation tensor is obtained as
\begin{equation}
P=P_{i}^{i}=-2\frac{\partial G^{i}}{\partial x^{i}}-2G^{l}G_{il}^{i}+ y^{l}\frac{\partial N_{i}^{i}}{\partial x^{l}}+N_{l}^{i}N_{i}^{l}+\frac{\partial N_{i}^{i}}{\partial t}.
\end{equation}

The third, fourth and fifth invariants of the system (\ref{EM}) are defined as \cite{An00}
\begin{equation}  \label{31}
P_{jk}^{i}\equiv \frac{1}{3}\left(\frac{\partial P_{j}^{i}}{\partial y^{k}} - \frac{\partial P_{k}^{i}}{\partial y^{j}}\right), \quad
P_{jkl}^{i}\equiv \frac{\partial P_{jk}^{i}}{\partial y^{l}}, \quad
D_{jkl}^{i}\equiv \frac{\partial G_{jk}^{i}}{\partial y^{l}}.
\end{equation}
The third invariant $P_{jk}^{i}$ can be interpreted geometrically as a torsion tensor. The fourth and fifth invariants $P_{jkl}^{i}$ and $D_{jkl}^{i}$ are called the Riemann-Christoffel curvature tensor, and the Douglas tensor, respectively \cite{An00,rev}. In a Berwald space, these tensors always exist. In the KCC theory, they describe the geometrical properties and interpretation of a system of second-order differential equations.

Alternatively, we can introduce another definition for the third and fourth KCC invariants \cite{T0}. In this formulation, the third KCC invariant is given by
\begin{equation}  \label{32}
B_{jk}^i=\frac{\delta N_j^i}{\delta x^k}-\frac{\delta N_k^i}{\delta x^j},
\end{equation}
where
\begin{equation}
\frac{\delta }{\delta x^i}=\frac{\partial}{\partial x^i}-N_i^j\frac{\partial}{\partial y^j},
\end{equation}
and the fourth KCC invariant can then be defined as
\begin{equation}
B_{jkl}^i=\frac{\partial B_{kl}^i}{\partial y^j}.
\end{equation}

In many physical, chemical and biological applications, we are interested in the behavior of the trajectories of the dynamical system (\ref{EM0}) in the vicinity of a point $x^{i}\left( t_{0}\right) $. For simplicity, in the following we take $t_{0}=0$. We consider the trajectories $x^{i}=x^{i}(t)$ as curves in the Euclidean space $\left(\mathbb{R}^{n},\left\langle .,.\right\rangle \right) $, where $\left\langle .,.\right\rangle $ is the canonical inner product of $\mathbb{R}^{n}$. For the deviation vector $\vec{\xi}$, we assume that it obeys the initial conditions $\vec{\xi} \left(0\right) = \vec{O}$ and $\dot{\vec{\xi}}\left(0\right) = \vec{W} \neq \vec{O}$, where $\vec{O} \in \mathbb{R}^{n}$ is the null vector \cite{An00, rev,Sa05,Sa05a}.

Thus, we introduce the following description of the focusing tendency of the trajectories around $t_{0}=0$: if $\left| \left| \xi \left( t\right) \right| \right| <t^{2}$, $t\approx 0^{+}$, the trajectories are {\it bunching together}. Conversely, if $\left| \left| \xi \left( t\right) \right| \right| >t^{2}$, $t\approx 0^{+}$, the trajectories are {\it dispersing} \cite{An00,rev,Sa05,Sa05a}. The focusing tendency of the trajectories can be also characterized in terms of the {\it deviation curvature tensor} in the following way  \cite{An00, rev, Sa05,Sa05a}:

{\it The trajectories of the system of equations (\ref{EM}) are bunching together for $t\approx 0^{+}$ if and only if the real parts of the eigenvalues of the deviation tensor $P_{j}^{i}\left( 0\right) $ are strictly negative.}

Conversely, {\it the trajectories are dispersing if and only if the real part of the eigenvalues of $P_{j}^{i}\left( 0\right) $ are strictly positive}.

Based on the above considerations, we define the concept of the Jacobi stability for a dynamical system as follows \cite{An00,rev, Sa05,Sa05a}:

\textbf{Definition:} Lets us assume that, with respect to the norm $\left| \left| .\right| \right| $ induced by a positive definite inner product, the system of differential equations (\ref{EM}) satisfies the initial conditions
\be
\left| \left| x^{i}\left( t_{0}\right) - \tilde{x}^{i}\left( t_{0}\right) \right| \right| =0, \quad \left| \left| \dot{x}^{i}\left( t_{0}\right) -\tilde{x}^{i}\left( t_{0}\right) \right| \right| \neq 0.
\ee

Then:

(a)  the trajectories of Eq.~(\ref{EM}) are {\it Jacobi stable if and only if the real parts of all of the eigenvalues of the deviation tensor $P_{j}^{i}$ are everywhere strictly negative};

(b) {\it if  the real parts of the eigenvalues of the deviation tensor $P_{j}^{i}$ are not everywhere strictly negative, the trajectories are Jacobi unstable}.

\begin{centering}
\begin{figure*}[htp]
\centering
\includegraphics[height=6cm,width=12cm]{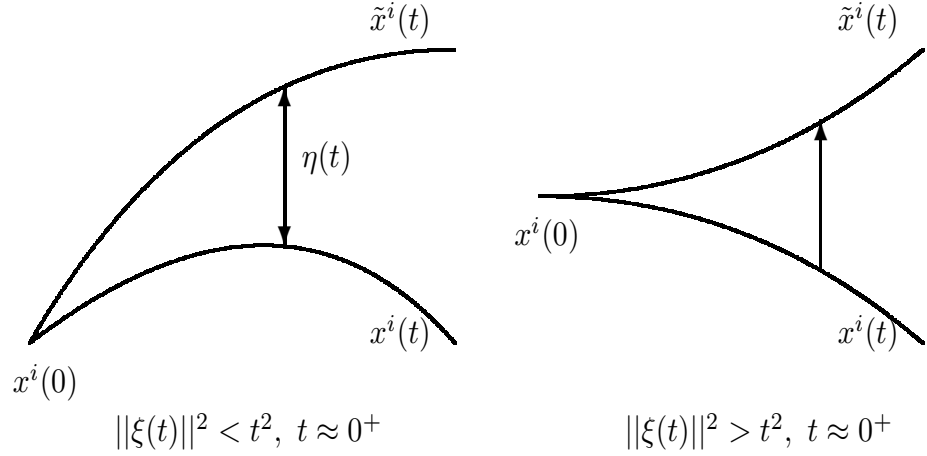}
\caption{Behavior of the trajectories near the initial configuration of the system.} \label{S4.1}
\label{pict1}
\end{figure*}
\end{centering}

Graphically, the focussing behavior of the trajectories near the origin is shown in Fig.~\ref{pict1}.

For a two-dimensional dynamical system, the components of the curvature deviation tensor can be written in a matrix form as
\begin{equation}
P_{j}^{i}=\left(
\begin{array}{c}
P_{1}^{1}\;\;\;\;P_{2}^{1} \\
P_{1}^{2}\;\;\;P_{2}^{2}
\end{array}
\right) ,
\end{equation}
and its eigenvalues are given by
\begin{equation}
\lambda _{\pm }=\frac{1}{2}\left[ P_{1}^{1}+P_{2}^{2}\pm \sqrt{\left(P_{1}^{1}-P_2^2\right)^{2}+4P_{2}^{1}P_{1}^{2}}\right],
\end{equation}
which are the roots of the characteristic quadratic equation
\begin{equation}  \label{eql}
\lambda ^2-\left(P_1^1+P_2^2\right)\lambda +\left(P_1^1P_2^2-P_2^1P_1^2\right) = 0.
\end{equation}

A very simple and efficient way to obtain the signs of the eigenvalues of the curvature deviation tensor is based on the Routh-Hurwitz criteria \cite{RH}. According to this criteria, all of the roots of the polynomial $P(\lambda)$ are negative, or have negative real parts, if the determinant of all Hurwitz matrices $\mathrm{det}\;H_j$, $j=1,2, \dots ,n$, are positive. For $n=2$, corresponding to the case of Eq.~(\ref{eql}), the Routh-Hurwitz criteria take the simple form
\begin{equation}
P_1^1+P_2^2<0,\quad P_1^1P_2^2-P_2^1P_1^2>0.
\end{equation}
Thus, $\lambda _{\pm}$ describe the curvature properties along a given geodesic. Another way to characterize the way in which the geodesics explore the geometry of the Finsler manifold is through the (half of the) Ricci curvature scalar along the flow, $\kappa$, and the anisotropy $\Theta $, defined as \cite{PR}
\begin{equation} \label{stability-1}
\kappa =\frac{1}{2}\left(\lambda _{+}+\lambda _{-}\right)=\frac{P}{2}=\frac{%
P_1^1+P_2^2}{2},
\end{equation}
and
\begin{equation} \label{stability-2}
\Theta =\frac{1}{2}\left(\lambda _{+}-\lambda _{-}\right)=\frac{\sqrt{\left(
P_{1}^{1}-P_2^2\right) ^{2}+4P_{2}^{1}P_{1}^{2} }}{2},
\end{equation}
respectively.

\section{Jacobi stability analysis of Model I} \label{S5}

In this Section, we use the KCC approach to study the dynamical properties of the string system proposed in Model I. We explicitly obtain the non-linear and Berwald connections, and the deviation curvature tensors. The eigenvalues of the deviation curvature tensor are also determined, and we study their properties at the equilibrium points of the wound string loop. Determining the signs of the eigenvalues allows us to formulate criteria for the Jacobi stability of the fixed points in terms of the wound string model parameters.

\subsection{The non-linear and Berwald connections, and the KCC invariants of Model I} \label{S5.1} 

By introducing the notation
\begin{equation}
\rho =X^{1}, \ \ \ R=X^{2}, \ \ \ \dot{\rho}=Y^{1}, \ \ \ \dot{R}=Y^{2},
\end{equation}
the EOM of the string system, Eqs.~(\ref{6}) and (\ref{7}), can be rewritten as
\begin{equation} \label{6a}
\frac{{\rm d}^{2}X^{1}}{{\rm d}t^{2}}+\frac{X^{1}}{a^{2}\left( X^{1}\right)^{2}
+ m^{2}\left( X^{2}\right)^{2}}\left[ a^{2}\left( 1-\left( Y^{1}\right)^{2}\right)
- \left( Y^{2}\right) ^{2}-C^{2}\frac{\left( a^{2}\left(X^{1}\right)^{2}
+ m^{2}\left( X^{2}\right) ^{2}\right) ^{2}}{\left(X^{1}\right) ^{4}\left( X^{2}\right) ^{2}}\right] =0,
\end{equation}
\begin{eqnarray} \label{7a}
\frac{{\rm d}^{2}x^{2}}{{\rm d}t^{2}} &+& \frac{x^{2}}{a^{2}\left( X^{1}\right)^{2}+m^{2}\left( X^{2}\right) ^{2}}
\nonumber\\
&\times& \left\{ m^{2}\left[ a^{2}\left(1-\left(Y^{1}\right) ^{2}\right) -\left(Y^{2}\right) ^{2}\right]
- a^{2}C^{2}\frac{\left(a^{2}\left( X^{1}\right)^{2}
+ m^{2}\left( X^{2}\right)^{2}\right)^{2}}{\left( X^{1}\right)^{2}\left( X^{2}\right)^{4}}\right\} = 0.
\end{eqnarray}
Equations (\ref{6a}) and (\ref{7a}) form a second order differential system, given by two equations of the form
\begin{equation}
\frac{d^{2}X^{i}}{dt^{2}}+2G^{i}\left( X^{i},Y^{i}\right) =0,\ i \in\left\{1,2\right\},
\end{equation}
where
\begin{equation}
G^{1} = \frac{1}{2}\frac{X^{1}}{a^{2}\left( X^{1}\right)^{2}
+ m^{2}\left( X^{2}\right)^{2}}\left[a^{2}\left( 1-\left( Y^{1}\right)^{2}\right)
- \left( Y^{2}\right)^{2}
- C^{2}\frac{\left( a^{2}\left(X^{1}\right) ^{2}+m^{2}\left(X^{2}\right)^{2}\right)^{2}}{\left( X^{1}\right)^{4}\left(X^{2}\right)^{2}}\right] ,
\end{equation}
\begin{equation}
G^{2} = \frac{1}{2}\frac{x^{2}}{a^{2}\left( X^{1}\right)^{2}
+ m^{2}\left( X^{2}\right) ^{2}}\left\{m^{2}\left[a^{2}\left(1-\left( Y^{1}\right) ^{2}\right)
- \left( Y^{2}\right)^{2}\right] - a^{2}C^{2}\frac{\left( a^{2}\left( X^{1}\right)^{2}
+ m^{2}\left( X^{2}\right) ^{2}\right) ^{2}}{\left( X^{1}\right)^{2}\left( X^{2}\right) ^{4}}\right\} .
\end{equation}

The coefficients of the non-linear connection associated to the system (\ref{6a}) and (\ref{7a}) are given by
\begin{equation}
N_{1}^{1} = \frac{\partial G^{1}}{\partial Y^{1}}=-\frac{a^2 X^1 Y^1}{a^2\left(X^1\right)^2+m^2 \left(X^2\right)^2},
\ \ \ N^{1}_{2}=\frac{\partial G^{1}}{\partial Y^{2}}=-\frac{X^1 Y^2}{a^2 \left(X^1\right)^2+m^2 \left(X^2\right)^2},
\end{equation}
\begin{equation}
N^{2}_{1} = \frac{\partial G^{2}}{\partial Y^{1}}=-\frac{a^2 m^2 X^1 Y^1}{a^2\left(X^1\right)^2+m^2 \left(X^2\right)^2},
\ \ \ N_{2}^{2}=\frac{\partial G^{2}}{\partial Y^{2}}=-\frac{m^2 X^1 Y^2}{a^2\left( X^1\right)^2+m^2\left(X^2\right)^2},
\end{equation}
and the components of the Berwald connection can be obtained as
\begin{equation}
G_{11}^1=\frac{\partial N_1^1}{\partial Y^1}=-\frac{a^2 X^1}{a^2\left(X^1\right)^2+m^2 \left(X^2\right)^2},
\ \ \ G_{12}^1=\frac{\partial N_1^1}{\partial Y^2}=0,
\end{equation}
\begin{equation}
G_{21}^1 = \frac{\partial N_2^1}{\partial Y^1}=0,
\ \ \ G_{22}^1 = \frac{\partial N_2^1}{\partial Y^2}=-\frac{X^1}{a^2 \left(X^1\right)^2+m^2 \left(X^2\right)^2},
\end{equation}
\begin{equation}
G_{11}^2 = \frac{\partial N^2_1}{\partial Y^1} = -\frac{a^2 m^2 X^1}{a^2\left(X^1\right)^2+m^2 \left(X^2\right)^2},
\ \ \ G^2_{12} = \frac{\partial N^2_1}{\partial Y^2}=0,
\end{equation}
\begin{equation}
G_{21}^2 = \frac{\partial N_2^2}{\partial Y^1}=0,
\ \ \ G_{22}^2 = \frac{\partial N_2^2}{\partial Y^2}=-\frac{m^2 X^1}{a^2 \left(X^1\right)^2+m^2 \left(X^2\right)^2}.
\end{equation}
The first KCC invariants are given by
\begin{equation}
\epsilon^{1}=\frac{a^{2}X^{1}}{a^{2}\left( X^{1}\right) ^{2}+m^{2}\left(X^{2}\right)^{2}}
- \frac{C^{2}\left[ a^{2}\left(X^{1}\right)^{2}+m^{2}\left( X^{2}\right) ^{2}\right] }{\left( X^{1}\right)^{3}\left(X^{2}\right) ^{2}},
\end{equation}
\begin{equation}
\epsilon^{2} = \frac{a^{2}m^{2}}{X^{1}}\left[ \frac{\left( X^{1}\right)^{2}}{a^{2}\left( X^{1}\right)^{2}
+ m^{2}\left( X^{2}\right)^{2}}-\frac{C^{2}}{\left(X^{2}\right)^{2}}\right]
- \frac{a^{4}C^{2}X^{1}}{\left( X^{2}\right)^{4}},
\end{equation}
and the components of the curvature deviation tensor are
\begin{eqnarray}
P_{1}^{1} &=& \frac{1}{\left(X^{1}\right)^{4}\left[ a^{2}\left(X^{1}\right)^{2}X_{2} + m^{2}\left( X^{2}\right)^{3}\right]^{2}}
\notag \\
&\times& \Bigg\{-2a^{6}C^{2}\left( X^{1}\right)^{6}+a^{4}\left(X^{1}\right)^{4}\left(X^{2}\right)^{2}\left( 2\left( X^{1}\right) ^{2}-7C^{2}m^{2}\right)
\notag \\
&+& \left[ -m^{2}\left(X^{2}\right) ^{2}\left( 8C^{2}m^{2}+\left( X^{1}\right) ^{2}\right)
+ m^{2}\left( X^{1}\right) ^{3}Y^{1}Y^{2}(X^{1}+2X^{2})-2\left( X^{1}\right)^{4}\left( Y^{2}\right) ^{2}\right]
\notag \\
&\times& a^{2}\left( X^{1}\right) ^{2}\left( X^{2}\right) ^{2}  - 3C^{2}m^{6}\left( X^{2}\right)^{6}
+ m^{2}\left( X^{1}\right) ^{4}\left(X^{2}\right) ^{4}\left( Y^{2}\right) ^{2}\Bigg\},
\end{eqnarray}
\begin{eqnarray}
P_{2}^{1} &=& \frac{1}{X^{1}\left(X^{2}\right)^{4}\left[a^{2}\left(X^{1}\right)^{2} + m^{2}\left( X^{2}\right)^{2}\right]^{2}}
\notag \\
&\times&
\Bigg\{X^{1}\left( X^{2}\right)^{4}Y^{1}Y^{2}\left[2a^{2}\left( X^{1}\right)^{2}-m^{2}\left(X^{2}\right) ^{2}\right] - a^{2}(X^{1}+2X^{2})
\notag\\
&\times&\left[a^{4}C^{2}\left( X^{1}\right) ^{4}+2a^{2}C^{2}m^{2}\left(X^{1}\right)^{2}\left( X^{2}\right)^{2}
+ m^{2}\left( X^{2}\right)^{4}\left( C^{2}m^{2}+\left( X^{1}\right)^{2}\left( \left( Y^{1}\right)^{2} -1\right) \right) \right]
\Bigg\},
\nonumber
\end{eqnarray}
\begin{eqnarray}
&&
\end{eqnarray}
\begin{eqnarray}
P_{1}^{2} &=& \frac{1}{\left( X^{2}\right) ^{4}\left[a^{2}\left(X^{1}\right)^{3} + m^{2}X^{1}\left( X^{2}\right) ^{2}\right]^{2}}
\notag \\
&\times&
\Bigg\{a^{8}C^{2}\left(X^{1}\right)^{6} + a^{4}m^{2}\left(X^{1}\right)^{2}\left(X^{2}\right) ^{4}\left[ 2\left( X^{1}\right) ^{2}-3C^{2}m^{2}\right]
\notag \\
&+& \left[ -m^{2}\left( X^{2}\right)^{2}\left( 2C^{2}m^{2}+\left( X^{1}\right)^{2}\right)
+ m^{2}\left(X^{1}\right) ^{3}Y^{1}Y^{2}(X^{1}+2X^{2})-2\left( X^{1}\right) ^{4}\left(Y^{2}\right) ^{2}\right]
\notag \\
&\times& a^{2}m^{2}\left(X^{2}\right)^{4}
+ m^{4}\left(X^{1}\right) ^{2}\left(X^{2}\right)^{6}\left(Y^{2}\right)^{2}\Bigg\},
\end{eqnarray}
\begin{eqnarray}
P_{2}^{2} &=& -\frac{1}{X^{1}\left( X^{2}\right)^{5}\left( a^{2}\left(X^{1}\right)^{2} + m^{2}\left( X^{2}\right) ^{2}\right)^{2}}
\notag \\
&\times&
\Bigg\{4a^{8}C^{2}\left( X^{1}\right)^{6}+a^{6}C^{2}m^{2}\left(X^{1}\right)^{4}X^{2}(X^{1}+10X^{2})
\notag\\
&+& \Bigg[m^{2}(X^{1}+2X^{2})\left(C^{2}m^{2}+\left(X^{1}\right)^{2}\left(\left(Y^{1}\right) ^{2}-1\right) \right) - 2\left( X^{1}\right) ^{3}Y^{1}Y^{2}\Bigg]
\notag\\
&\times& 2a^{4}C^{2}m^{4}\left( X^{1}\right)^{2}\left( X^{2}\right)^{3}(X^{1} + 4X^{2})+a^{2}m^{2}\left(X^{2}\right)^{5}
\notag\\
&+& m^{4}X^{1}\left(X^{2}\right)^{6}Y^{2}\left[ 2X^{1}(Y^{2}-1) + X^{2}Y^{1}\right] \Bigg\}.
\end{eqnarray}

\subsection{Jacobi stability analysis of the critical points of Model I} \label{S5.2} 

As the next step in the Jacobi stability analysis, we calculate the components of the curvature deviation tensor at the critical points of the system, $X^1 = \pm 2mC$, $X^2 = \pm 2aC$, $Y^1=0$, $Y^2=0$. Thus, we obtain
\begin{eqnarray}
P^1_1\left(X^1=\pm2mC,X^2=\pm2aC,Y^1=0,Y^2=0\right) &{}&
\nonumber\\
= P^1_1\left(X^1=\pm2mC,X^2=\mp2aC,Y^1=0,Y^2=0\right) &=& -\frac{1}{4 m^2C^2},
\nonumber\\
P^1_2\left(X^1=\pm2mC,X^2=\pm2aC,Y^1=0,Y^2=0\right) &{}&
\nonumber\\
= P^1_2\left(X^1=\pm2mC,X^2=\mp2aC,Y^1=0,Y^2=0\right) &=& 0,
\end{eqnarray}
\begin{eqnarray}
P^2_1\left(X^1=\pm2mC,X^2=\pm2aC,Y^1=0,Y^2=0\right) &{}&
\nonumber\\
= P^2_1\left(X^1=\pm2mC,X^2=\mp2aC,Y^1=0,Y^2=0\right) &=& 0,
\nonumber\\
P^2_2\left(X^1=\pm2mC,X^2=\pm2aC,Y^1=0,Y^2=0\right) &{}&
\nonumber\\
= -P^2_2\left(X^1=\pm2mC,X^2=\mp2aC,Y^1=0,Y^2=0\right) &=& -\frac{m}{4 a C^2}.
\end{eqnarray}

We therefore obtain the following theorem:

\textbf{Theorem 1} (a) {\it If the parameters $a$, $m$ and $C$ of the string Model I simultaneously satisfy the conditions
\begin{equation} \label{m_cond1}
-\frac{a + m^3}{4am^2C^2} < 0, \ \ \ \frac{1}{16amC^2} > 0
\end{equation}
then the critical points $X^1=\pm2mC$, $X^2=\pm2aC$, $Y^1=0$, $Y^2=0$ of the string system are Jacobi stable, and are Jacobi unstable otherwise.}

(b) {\it If the parameters $a$, $m$ and $C$ of the string Model I simultaneously satisfy the conditions
\begin{equation} \label{m_cond2}
\frac{-a + m^3}{4am^2C^2} < 0, \ \ \ -\frac{1}{16amC^2} > 0
\end{equation}
then the critical points $X^1=\pm2mC$, $X^2=\mp2aC$, $Y^1=0$, $Y^2=0$ of the string system are Jacobi stable, and are Jacobi unstable otherwise.}

\textbf{Corollary.} {\it The stability conditions of Theorem 1(a) and (b) reduce to the individually equivalent conditions
\begin{equation} \label{cor1a}
am > 0, \ \ \ am < 0.
\end{equation}}

The conditions in Eq. (\ref{cor1a}) appear to indicate that, in flux-compactified geometries, the stability of different winding states is determined by the warp factor induced by the back reaction on the large dimensions. Mathematically, the sign of $a$ appears to be arbitrary, since only $a^2$ appears in the metric of the background space-time. However, in general, we may take a factor of $a^2$ outside the denominator $\sqrt{-\gamma}$ in the expression for the Hamiltonian (see, for example, Eq. (\ref{E-l})), so that the positivity of the energy requires us to take the positive square root. On physical grounds therefore, we may set $a>0$. By convention, we may also require $C$ to have the same sign as $m$. This is in accordance with the identification $C=l/E$, where we choose positive angular momentum to be associated with $m>0$ and negative angular momentum to be associated with $m <0$ (c.f. Eq. (\ref{E-l})).

Hence, if $m>0$, only one critical point from the first set (Theorem 1(a)) is physically relevant, $X^1=+2mC$, $X^2=+2aC$, whereas neither of the critical points from the second set (Theorem 1(b)) are physically valid. In this case, the single remaining physical critical point is Jacobi stable, by the first condition in the Corollary, Eq. (\ref{cor1a}). If $m<0$, neither of the critical points belonging to the first set is physical, and only one point from the second set remains: $X^1=2mC$, $X^2=-2aC$. This is Jacobi stable by the second condition in the Corollary, Eq. (\ref{cor1a}). In this case, $C$ is negative, so that the stable critical point for $m<0$ is, in fact, completely equivalent to the stable critical point for $m>0$, as expected by symmetry. At all critical points, all components of the first KCC invariant are zero.

These results are of interest to all cosmic string models in string theory inspired models of the early universe, including brane inflation scenarios in which copious $F$ and $D-$string production occurs \cite{Sarangi:2002yt,Jones:2003da,Pogosian:2003mz}. In such models, the compact dimensions are flux compactified, which leads, generically, to the existence of a warp factor due to the back reaction on the non-compact space.

\subsection{The behavior of the deviation vector in Model I}
\label{S5.3}

The time evolution of the deviation vector $\xi ^{i}$, $i \in \left\{1,2\right\}$, giving the behavior of the trajectories of a dynamical system near a fixed point $x^{i}\left(t_{0}\right) $, is described by Eqs.~(\ref{def}) and (\ref{JE}). In the case of the string system proposed in Model I, these equations can be written as
\begin{eqnarray}
&&\frac{{\rm d}^{2}\xi^{1}(t)}{{\rm d}t^{2}}-\frac{2a^{2}X^{1}Y^{1}}{a^{2}\left( X^{1}\right)^{2}
+ m^{2}\left( X^{2}\right) ^{2}}\frac{{\rm d}\xi^{1}(t)}{{\rm d}t}
- \frac{2X^{1}Y^{2}}{a^{2}\left( X^{1}\right) ^{2}+m^{2}\left(X^{2}\right)^{2}}\frac{{\rm d}\xi^{2}(t)}{{\rm d}t}
\notag \\
&+&\frac{1}{X^{1}\left( X^{2}\right)^{3}\left(a^{2}\left( X^{1}\right)^{2} + m^{2}\left( X^{2}\right)^{2}\right) ^{2}}
\notag  \label{56} \\
&\times &\Bigg\{ 2\Big(a^{6}C^{2}\left(X^{1}\right)^{4}+2a^{4}C^{2}m^{2}\left( X^{1}\right)^{2}\left( X^{2}\right)^{2}
+ a^{2}m^{2}\left( X^{2}\right) ^{4}\left( C^{2}m^{2}+\left( X^{1}\right)^{2}\left(\left(Y^{1}\right)^{2}-1\right) \right)
\notag \\
&+&m^{2}\left(X^{1}\right)^{2}\left(X^{2}\right)^{4}\left(Y^{2}\right)^{2}\Big)\Bigg\} \text{$\xi^{2}$}(t)
+ \frac{1}{\left(X^{1}\right)^{4}\left(X^{2}\right)^{2}\left(a^{2}\left(X^{1}\right)^{2} + m^{2}\left(X^{2}\right) ^{2}\right)^{2}}
\notag \\
&\times& \Bigg\{a^{6}C^{2}\left( X^{1}\right)^{6}
+ a^{4}\left(X^{1}\right)^{4}\left(X^{2}\right)^{2}\left[5C^{2}m^{2} + \left(X^{1}\right)^{2}\left(\left(Y^{1}\right) ^{2}-1\right)\right]
\notag \\
&+& a^{2}\left(X^{1}\right)^{2}\left(X^{2}\right)^{2}\left[m^{2}\left(X^{2}\right)^{2}\left(7C^{2}m^{2}
- \left( X^{1}\right)^{2}\left(\left(Y^{1}\right) ^{2}-1\right) \right) + \left(X^{1}\right)^{4}\left(Y^{2}\right)^{2}\right]
\notag \\
&+& 3C^{2}m^{6}\left( X^{2}\right) ^{6} - m^{2}\left( X^{1}\right)^{4}\left(X^{2}\right)^{4}\left(Y^{2}\right) ^{2}\Bigg\}\xi^{1}(t) = 0,
\end{eqnarray}
\begin{eqnarray}
&& \frac{{\rm d}^{2}\xi^{2}(t)}{{\rm d}t^{2}} - \frac{2a^{2}m^{2}X^{1}Y^{1}}{a^{2}\left(X^{1}\right)^{2} + m^{2}\left(X^{2}\right)^{2}}\frac{{\rm d}\xi^{1}(t)}{{\rm d}t}
- \frac{2m^{2}X^{1}Y^{2}}{a^{2}\left(X^{1}\right)^{2} + m^{2}\left(X^{2}\right) ^{2}}\frac{{\rm d}\xi^{2}(t)}{{\rm d}t}
\notag  \label{57} \\
&-& \frac{1}{\left(X^{1}\right)^{2}\left(X^{2}\right)^{4}\left(a^{2}\left(X^{1}\right)^{2} + m^{2}\left(X^{2}\right) ^{2}\right)^{2}}
\notag \\
&\times& \Bigg\{(aX^{1}-mX^{2})(aX^{1}+mX^{2})a^{6}C^{2}\left(X^{1}\right)^{4} + 2a^{4}C^{2}m^{2}\left(X^{1}\right)^{2}\left(X^{2}\right)^{2}
\notag \\
&+& a^{2}m^{2}\left( X^{2}\right)^{4}\left[ C^{2}m^{2} - \left( X^{1}\right)^{2}\left(\left(Y^{1}\right)^{2}-1\right) \right] - m^{2}\left(X^{1}\right) ^{2}\left(X^{2}\right)^{4}\left(Y^{2}\right)^{2}\Bigg\}\xi^{1}(t)
\notag \\
&+&\frac{1}{X^{1}\left( X^{2}\right)^{5}\left(a^{2}\left(X^{1}\right)^{2} + m^{2}\left(X^{2}\right)^{2}\right)^{2}}
\notag \\
&\times& \Bigg\{4a^{8}C^{2}\left(X^{1}\right)^{6}+5a^{6}C^{2}m^{2}\left(X^{1}\right)^{4}\left(X^{2}\right)^{2}
+ 4a^{4}C^{2}m^{4}\left(X^{1}\right)^{2}\left(X^{2}\right)^{4}
\notag \\
&+& a^{2}m^{4}\left(X^{2}\right)^{6}\left[C^{2}m^{2} + \left(X^{1}\right)^{2}\left(\left(Y^{1}\right)^{2} - 1\right)\right]  + m^{4}\left(X^{1}\right)^{2}\left(X^{2}\right)^{6}\left(Y^{2}\right)^{2}\Bigg\}\xi^{2}(t)=0.
\end{eqnarray}

By evaluating Eqs.~(\ref{56}) and (\ref{57}) at $X^{1}=\pm2mC $, $X^{2}=\pm2aC$, $Y^{1}=0$, $Y^{2}=0$, we obtain the evolution equations of the deviation vector near the critical points as
\begin{equation} \label{dev-1}
\frac{{\rm d}^{2}\xi^{1}(t)}{{\rm d}t^{2}} + \frac{4}{C^{2}m^{2}}\xi^{1}(t)=0,
\end{equation}
\begin{equation} \label{dev-2}
\frac{{\rm d}^{2}\xi ^{2}(t)}{{\rm d}t^{2}} \pm \frac{m}{4aC^{2}}\xi^{2}(t)=0,
\end{equation}
where we must take the positive sign in front of the second term in Eq. (\ref{57}) if the choice of signs in the expressions for $X^1$ and $X^2$ match, and the negative sign if they are opposite. In the first scenario ($am>0$), the general solutions to these equations are given by
\begin{eqnarray}
\xi^{1}(t) = 2mC\dot{\xi}^{1}_{0}\sin\left(\frac{1}{2mC}\ t\right), 
\end{eqnarray}
and
\begin{equation}
\xi^{2}(t) = 2C\sqrt{\frac{a}{m}}\dot{\xi}^{2}_{0}\sin\left(\frac{1}{2C}\sqrt{\frac{m}{a}}\ t\right), 
\end{equation}
respectively, where we have used the initial conditions $\xi ^{1}(0)=0$, $\xi ^{2}(0)=0$, ${\rm d}\xi ^{1}(t)/{\rm d}t|_{t=0}=\dot{\xi}^{1}_{0}$, ${\rm d}\xi^{2}(t)/{\rm d}t|_{t=0}=\dot{\xi}^{2}_{0}$. In the second case ($am<0$), the general solution to Eq. (\ref{dev-2}) is
\begin{equation} \label{dev-3}
\xi ^{2}(t) = C\sqrt{\frac{a}{m}}\dot{\xi}^{2}_{0}\left[\exp\left(\frac{1}{2C}\sqrt{\frac{m}{a}}\ t\right)
- \exp\left(-\frac{1}{2C}\sqrt{\frac{m}{a}}\ t\right)\right]=2C\sqrt{\frac{a}{m}}\dot{\xi}^{2}_{0}\sinh \left(\frac{1}{2C}\sqrt{\frac{m}{a}}\ t\right). 
\end{equation}
Since both $a/m<0$ and $m/a<0$, with the use of the relation $\sinh (ix)=i\sin x$, it immediately follows that Eq.~(\ref{dev-3}) can be written as
\be
\xi ^2(t)=-2C\sqrt{\left|\frac{a}{m}\right|}\dot{\xi}^{2}_{0}\sin \left(\frac{1}{2C}\sqrt{\left|\frac{m}{a}\right|}\ t\right),
\ee
thus showing that the time evolution of the deviation vector is bounded. Therefore, for $am<0$, the critical points are also Jacobi stable.

\section{Jacobi stability analysis of Model II} \label{S6}

In this Section, we use the KCC approach to study the dynamical properties of the string system proposed in Model II, following the same basic procedure as in Sect. \ref{S5}.

\subsection{The nonlinear and Berwald connections, and the KCC invariants of Model II} \label{S6.1}

By introducing the notation
\begin{equation}
\rho =X^1, \ \ \ \theta =X^2, \ \ \ \dot{\rho}=Y^1, \ \ \ \dot{\theta }=Y^2,
\end{equation}
we can write Eqs. (\ref{s21}) and (\ref{s22}) in the form of Eq.~(\ref{EM}), where
\begin{equation}
G^1 = \frac{1}{2} X^1 \left\{\frac{a^2 \left[1-\left(Y^1\right)^2\right]-\mathcal{R}^2 \left(Y^2\right)^2}{a^2\left(X^1\right)^2+m^2 \mathcal{R}^2 \sin^2(X^2)}
- \frac{\csc ^2(X^2) \left[a^2 \left(X^1\right)^2+m^2 \sin ^2(X^2)\right]}{\mathcal{R}^2\left(X^1\right)^4}\right\},
\end{equation}
\begin{eqnarray} \label{G2}
&&G^2 = \frac{1}{2} \sin (X^2) \cos (X^2)
\notag \\
&\times&\left\{\frac{m^2 \left[a^2 \left(1-\left(Y^1\right)^2\right)-\mathcal{R}^2\left(Y^2\right)^2\right]}{a^2 \left(X^1\right)^2+m^2 \mathcal{R}^2 \sin ^2(X^2)}
- \frac{\csc ^4(X^2) \left[a^2 \left(X^1\right)^2+m^2 \mathcal{R}^2 \sin ^2(X^2)\right]}{\mathcal{R}^2 \left(X^1\right)^2}\right\}.
\end{eqnarray}
The components of the nonlinear connection $N^i_j$ are given by
\begin{equation}
N_1^1 = -\frac{a^2 X^1 Y^1}{a^2\left(X^1\right)^2+m^2 \mathcal{R}^2 \sin ^2(X^2)}, \ \ \
N_1^2 = -\frac{\mathcal{R}^2 \left(X^1\right)Y^2}{a^2\left(X^1\right)^2+m^2 \mathcal{R}^2 \sin ^2(X^2)},
\end{equation}
\begin{equation}
N_2^1 = -\frac{a^2 m^2 Y^1 \sin (X^2) \cos (X^2)}{a^2 \left(X^1\right)^2+m^2 \mathcal{R}^2 \sin ^2(X^2)}, \ \ \
N_2^2 = -\frac{m^2 \mathcal{R}^2 Y^2 \sin (X^2) \cos (X^2)}{a^2 \left(X^1\right)^2+m^2 \mathcal{R}^2 \sin ^2(X^2)},
\end{equation}
and the components of the Berwald connection are
\begin{equation}
G^1_{11} = -\frac{a^2 X^1}{a^2\left(X^1\right)^2+m^2 \mathcal{R}^2 \sin ^2(X^2)}, \ \ \
G^1_{12} = 0, \ \ \
G^1_{21} = 0,
\end{equation}
\begin{equation}
G^1_{22} = -\frac{R_0^2 X^1}{a^2\left(X^1\right)^2+m^2 \mathcal{R}^2 \sin ^2(X^2)}, \ \ \
G^2_{11} = -\frac{a^2 m^2 \sin (X^2) \cos (X^2)}{a^2 \left(X^1\right)^2+m^2\mathcal{R}^2 \sin ^2(X^2)},
\end{equation}
\begin{equation}
G^2_{12} = 0, \ \ \
G^2_{21} = 0, \ \ \
G^2_{22} = -\frac{m^2 \mathcal{R}^2 \sin (X^2) \cos (X^2)}{a^2\left(X^1\right)^2+m^2 \mathcal{R}^2 \sin ^2(X^2)}.
\end{equation}
The components of the first KCC invariant can then be obtained as
\begin{equation}
\epsilon^1 = -\frac{m^2 C^2}{\left(X^1\right)^3}-\frac{a^2C^2 \csc ^2(X^2)}{\mathcal{R}^2 X^1}+\frac{2 a^2 X^1}{2 a^2 \left(X^1\right)^2-m^2 \mathcal{R}^2 \cos (2 X^2)+m^2 \mathcal{R}^2},
\end{equation}
\begin{equation}
\epsilon^2 = -\frac{a^2}{\mathcal{R}^2C^2}\cot (X^2) \left[\frac{m^2}{\left(X^1\right)^2}
+ \frac{a^2}{\mathcal{R}^2}\csc ^2(X^2)\right]-\frac{a^2 m^2 \sin(2X^2)}{2 a^2 \left(X^1\right)^2 - m^2 \mathcal{R}^2 \cos (2 X^2) + m^2\mathcal{R}^2},
\end{equation}
and the components of the curvature deviation tensor are
\begin{eqnarray}
P_{1}^{1} &=& -\frac{3m^{2}C^2}{\left( X^{1}\right)^{4}}-\frac{2a^2C^2\csc^{2}(X^{2})}{\mathcal{R}^2\left( X^{1}\right)^{2}}
- \frac{2(a-\mathcal{R}Y^{2})(a+\mathcal{R}Y^{2})}{2a^{2}\left( X^{1}\right)^{2} - m^{2}\mathcal{R}^{2}\cos (2X^{2})+m^{2}\mathcal{R}^{2}}
\notag \\
&+&\frac{2a^{2}X^{1}\left[6X^{1}(a-\mathcal{R}Y^{2})(a+\mathcal{R}Y^{2})
+ m^{2}\mathcal{R}^{2}Y^{1}Y^{2}\sin (2X^{2})\right] }{\left[ 2a^{2}\left( X^{1}\right)^{2} - m^{2}\mathcal{R}^{2}\cos(2X^{2})+m^{2}\mathcal{R}^{2}\right]^{2}}
\end{eqnarray}
\begin{eqnarray}
P_{2}^{1} &=& -\frac{a^2C^2}{\mathcal{R}^2X^1[a^2(X^1)^2 + m^2\mathcal{R}^2\sin^2(X^2)]}\Bigg\{\frac{m^{2}}{2}\sin (2X^{2})
\notag \\
&\times& \left[\frac{\mathcal{R}^4}{C^2}\left( X^{1}\right) ^{2}\left( \left( Y^{1}\right) ^{2}-1\right)+m^{2}\mathcal{R}^{4}\right]
+ \frac{\mathcal{R}^4}{a^2C^2}X^{1}Y^{1}Y^{2}\left[ m^{2}\mathcal{R}^{2}\sin^{2}(X^{2})-2a^{2}\left( X^{1}\right) ^{2}\right]
\notag \\
&+& \cot (X^{2})\left[a^{4}\left( X^{1}\right) ^{4}\csc^{2}(X^{2})+2a^{2}m^{2}\mathcal{R}^{2}\left(X^{1}\right) ^{2}\right] \Bigg\},
\end{eqnarray}
\begin{eqnarray}
P_{1}^{2} &=& \frac{3m^{2}a^2C^2\cot(X^{2})}{\mathcal{R}^{2}a^{2}C^{2}\left( X^{1}\right) ^{3}}-\frac{a^{2}Y^{1}Y^{2}}{\mathcal{R}^{2}}
\notag \\
&+&\frac{4a^{4}\left( X^{1}\right) ^{2}Y^{1}Y^{2}\left[ a^{2}\left(X^{1}\right)^{2} + m^{2}\mathcal{R}^{2}\right] -6a^{2}m^{2}\mathcal{R}^{2}X^{1}\sin(2X^{2})(a-\mathcal{R}Y^{2})(a+\mathcal{R}Y^{2})}{\mathcal{R}^{2}\left[2a^{2}\left(X^{1}\right)^{2} - m^{2}\mathcal{R}^{2}\cos (2X^{2}) + m^{2}\mathcal{R}^{2}\right]^{2}},
\nonumber
\end{eqnarray}
\begin{eqnarray}
{}
\end{eqnarray}
\begin{eqnarray}
P_{2}^{2} &=& \frac{a^2C^2}{\mathcal{R}^4\left(X^{1}\right)^{2}\left[ a^{2}\left(X^{1}\right)^{2}
+ m^{2}\mathcal{R}^{2}\sin^{2}(X^{2})\right]^{2}} \Bigg\{a^{6}\left( X^{1}\right) ^{6}(\cos(2X^{2})+2)\csc^{4}(X^{2})
\notag \\
&-& 3m^{2}\frac{\mathcal{R}^{6}}{C^2}\left(X^{1}\right)^{3}Y^{1}Y^{2}\sin(X^{2})\cos(X^{2}) + a^{2}m^{2}\left(X^{1}\right) ^{2}\sin^{2}(X^{2})
\notag \\
&\times& \Bigg[\frac{\mathcal{R}^4}{a^{2}C^{2}}\left(X^{1}\right)^{2}\left(a^{2}\left(\left(Y^{1}\right)^{2}-1\right)
+ \mathcal{R}^{2}(Y^{2}-1)Y^{2}\right) + m^{2}\frac{\mathcal{R}^8}{a^4C^2}(Y^{2}-1)Y^{2}\cos ^{2}(X^{2})  \notag \\
&&+3m^{2}\mathcal{R}^{4}\Bigg]
+ a^{2}m^{2}\left(X^{1}\right) ^{2}\Bigg[\cos^{2}(X^{2})\Bigg(\frac{\mathcal{R}^4}{a^{2}C^{2}}\left(X^{1}\right)^{2}\Bigg(a^{2}\left(-\left( \left(Y^{1}\right) ^{2}-1\right) \right)
 \notag \\
&&-\mathcal{R}^{2}\left(Y^{2}-1\right)Y^{2}\Bigg) + 3m^{2}\mathcal{R}^{4}\Bigg)
+ 3 a^{2}\mathcal{R}^{2}\left( X^{1}\right) ^{2}\left(2\cot ^{2}(X^{2}) + 1\right)\Bigg] +m^{4}\mathcal{R}^{2}\sin ^{4}(X^{2})
\notag \\
&\times& \Bigg[\frac{\mathcal{R}^4}{a^{2}C^{2}}\left(X^{1}\right)^{2}  \left(a^{2}\left(\left(Y^{1}\right)^{2}-1\right)
+ \mathcal{R}^{2}(Y^{2}-1)Y^{2}\right) + m^{2}\mathcal{R}^{4}\Bigg] \Bigg\}.
\end{eqnarray}

\subsection{Jacobi stability analysis of the critical points of Model II} \label{S6.2}

At the first set of critical points, for which $X^2 \neq \pi/2$, ($C \neq \pm (1/2)\mathcal{R}/a$), we obtain the following expressions for the components of the deviation curvature tensor:
\begin{eqnarray} \label{Eval-1}
P^1_1\left(X^1 = \pm2mC,X^2 = \pm \sin^{-1}(2aC/\mathcal{R}),0,0\right) &{}&
\nonumber\\
= P^1_1\left(X^1 = \pm2mC,X^2 = \mp\sin^{-1}(2aC/\mathcal{R}),0,0\right) &=& -\frac{1}{4m^2C^2},
\nonumber\\
P^1_2\left(X^1 = \pm2mC,X^2 = \pm \sin^{-1}(2aC/\mathcal{R}),0,0\right) &{}&
\nonumber\\
= P^1_2\left(X^1 = \pm2mC,X^2 = \mp \sin^{-1}(2aC/\mathcal{R}),0,0\right) &=& 0,
\nonumber\\
P^2_1\left(X^1 = \pm2mC,X^2 = \pm \sin^{-1}(2aC/\mathcal{R}),0,0\right) &{}&
\nonumber\\
= P^2_1\left(X^1 = \pm2mC,X^2 = \mp \sin^{-1}(2aC/\mathcal{R}),0,0\right) &=& 0,
\nonumber\\
P^2_2\left(X^1 = \pm2mC,X^2 = \pm \sin^{-1}(2aC/\mathcal{R}),0,0\right) &{}&
\nonumber\\
= P^2_2\left(X^1 = \pm2mC,X^2 = \mp \sin^{-1}(2aC/\mathcal{R}),0,0\right) &=& \frac{a^2}{\mathcal{R}^2} - \frac{1}{4C^2}.
\end{eqnarray}

At the first second set of critical points, for which $X^2 = \pi/2$, ($C = \pm (1/2)\mathcal{R}/a$), we obtain, instead:
\begin{eqnarray} \label{Eval-2}
&&P^1_1\left(X^1 = \pm \sqrt{\frac{C}{\pm \mathcal{R} - aC}}\frac{m\mathcal{R}}{\sqrt{a}},X^2 = \pi/2,0,0\right) = -\frac{4a^2(\mp aC+R)^2}{m^2\mathcal{R}^2},
\nonumber\\
&&P^1_2\left(X^1 = \pm \sqrt{\frac{C}{\pm \mathcal{R} - aC}}\frac{m\mathcal{R}}{\sqrt{a}},X^2 = \pi/2,0,0\right) = 0
\nonumber\\
&&P^2_1\left(X^1 = \pm \sqrt{\frac{C}{\pm \mathcal{R} - aC}}\frac{m\mathcal{R}}{\sqrt{a}},X^2 = \pi/2,0,0\right) = 0
\nonumber\\
&&P^2_2\left(X^1 = \pm \sqrt{\frac{C}{\pm \mathcal{R} - aC}}\frac{m\mathcal{R}}{\sqrt{a}},X^2 = \pi/2,0,0\right) = 0
\end{eqnarray}

From the general criteria for Jacobi stability, Eqs. (\ref{stability-1}) and (\ref{stability-2}), we therefore obtain the following theorem:

\textbf{Theorem 2} (a) {\it If the parameters $a$, $m$, $C$ and $\mathcal{R}$ of the string Model II simultaneously satisfy the conditions
\begin{equation} \label{Thm2}
-\frac{m^2+1}{4m^2C^2} + \frac{a^2}{\mathcal{R}^2} < 0, \ \ \ \frac{\mathcal{R}^2-4a^2C^2}{16m^2C^4\mathcal{R}^2} > 0,
\end{equation}
then the critical points $X^{1}=\pm2mC$, $X^{2} = \pm \sin^{-1}(2aC/\mathcal{R})$, $Y^1=0$, $Y^2=0$ and $X^{1}=\pm2mC$, $X^{2}$ $= \mp \sin^{-1}(2aC/\mathcal{R})$, $(C \neq \pm (1/2)\mathcal{R}/a)$, $Y^1=0$, $Y^2=0$ of the string system are Jacobi stable, and are Jacobi unstable otherwise.}

(b) {\it If the parameters $a$, $m$, $C$ and $\mathcal{R}$ of the string Model II satisfy the conditions
\begin{equation} \label{Thm2}
-\frac{4a^2(\mp aC+R)^2}{m^2\mathcal{R}^4} < 0,
\end{equation}
then the critical points $X^{1}=\pm \sqrt{C/(\pm \mathcal{R} - aC)}(m\mathcal{R}/\sqrt{a})$, $X^{2} = \pi/2$, of the string system are Jacobi stable, and are Jacobi unstable otherwise.}

\textbf{Corollary} {\it The stability conditions of Theorem 2(a) may be rewritten as
\begin{equation} \label{Thm2*}
C^2 < \frac{1}{4}\frac{(m^2+1)}{m^2}\frac{\mathcal{R}^2}{a^2}, \ \ \ C^2 < \frac{1}{4}\frac{\mathcal{R}^2}{a^2}.
\end{equation}
and those of Theorem 2(b) are equivalent to
\begin{equation} \label{Thm2**}
C^2 <  \mathcal{R}^2/a^2.
\end{equation}}

It is clear that if the second condition in Eq. (\ref{Thm2*}) is satisfied, the first is automatically satisfied as well. Furthermore, the first condition is equivalent to $\sin^2\theta < (m^2+1)/m^2$, which is automatically satisfied for any value of the parameter $m^2\geq1$, while the second is equivalent to $\sin^2\theta < 1$, which is automatically satisfied for $\theta < \pi/2$.

The condition (\ref{Thm2**}) is simply equivalent to the positivity condition on $X^1 = \rho$, obtained in Sect. \ref{S3.4}. This is also automatically satisfied, if we require continuity of the critical points of the system in the limit $C^2 \rightarrow (1/4)\mathcal{R}^2/a^2$, which strongly suggested by comparing the expressions for $P^{2}_{2}$ obtained from Eqs. (\ref{Eval-1}) and (\ref{Eval-2}). Substituting $C = \pm (1/2)\mathcal{R}/a$, explicitly, into Eq.  (\ref{Thm2}), the conditions become trivial (i.e., $-a^2/(m^2\mathcal{R}^2)<0$). Hence, we see that, even for $C^2 = (1/4)\mathcal{R}^2/a^2$, the system remains Jacobi stable. In this case, the second order EOM in $\theta(t)$, Eq. (\ref{s22}), reduces to $\ddot{\theta} = 0$ (identically) and $G^2 = 0$ (c.f. Eq. (\ref{G2}) in this limit), but our physical results are continuous with those obtained for $C^2 < (1/4)\mathcal{R}^2/a^2$.

Finally, we also note that the two bounds in Eq. (\ref{Thm2*}) coincide in the limit $m^2 \rightarrow \infty$, yielding $C^2 < (1/4)\mathcal{R}^2/a^2$ as the single remaining condition. Since the inequality is strict, the upper limit of this bound, $C^2 \rightarrow (1/4)\mathcal{R}^2/a^2$, which corresponds to the $\rho = const.$, $\Omega^2=1/2$, $\theta = \pi/2$ solution, in which the windings wrap great circles in the $S^2$, must be included in a separate analysis. When evaluated, this effectively recovers the critical points of Model I. Thus, for very high angular momenta, the winding radius of the string is dynamically driven to its maximum value, $\theta \rightarrow \pi/2$. As in Model I, the components of the first KCC invariant vanishes identically at all critical points.

\subsection{The behavior of the deviation vector in Model II} \label{S6.3} 

We now consider the time evolution of the components $\xi ^{i}$, $i \in \left\{1,2\right\}$, of the deviation vector, giving the dynamical behavior of the system trajectories near a fixed point $x^{i}\left(t_{0}\right)$ in Model II. The behavior of the deviation vector is described by Eqs.~(\ref{def}) and (\ref{JE}), which, for the second string model, take the form
\begin{eqnarray} \label{56*}
&& \frac{{\rm d}^2\xi^{1}(t)}{{\rm d}t^2} - \frac{4a^{2}X^{1}Y^{1}}{2a^{2}\left(X^{1}\right) ^{2}-m^{2}\mathcal{R}^{2}\cos (2X^{2})+m^{2}\mathcal{R}^{2}}\frac{{\rm d}\xi^{1}(t)}{{\rm d}t}
\notag \\
&+& \frac{4\mathcal{R}^{2}X^{1}Y^{2}}{-2a^{2}\left(X^{1}\right)^{2}+m^{2}\mathcal{R}^{2}\cos (2X^{2})-m^{2}\mathcal{R}^{2}}\frac{{\rm d}\xi^{2}(t)}{{\rm d}t}
+ \Bigg[ \frac{3m^{2}C^2}{\left( X^{1}\right)^{4}}+\frac{a^2C^2\csc ^{2}(X^{2})}{\mathcal{R}^2\left( X^{1}\right)^{2}}
\notag \\
&+& \frac{2\left(a^{2}\left(\left(Y^{1}\right) ^{2}-1\right)
+ \mathcal{R}^{2}\left(Y^{2}\right)^{2}\right) }{-2a^{2}\left( X^{1}\right)^{2}+m^{2}\mathcal{R}^{2}\cos (2X^{2}) -m^{2}\mathcal{R}^{2}}
+ \frac{8a^{2}\left(X^{1}\right)^{2}\left( a^{2}\left( \left( Y^{1}\right) ^{2}-1\right)
+ \mathcal{R}^{2}\left( Y^{2}\right) ^{2}\right) }{\left( 2a^{2}\left( X^{1}\right)^{2}-m^{2}\mathcal{R}^{2}\cos (2X^{2})+m^{2}\mathcal{R}^{2}\right) ^{2}}\Bigg] \xi^{1}(t)
\notag \\
&+& \left[ \frac{2a^2C^2\cot (X^{2})\csc ^{2}(X^{2})}{\mathcal{R}^2X^{1}}+\frac{4m^{2}\mathcal{R}^{2}X^{1}\sin(2X^{2})\left(a^{2}\left(\left(Y^{1}\right)^{2}-1\right)
+ \mathcal{R}^{2}\left(Y^{2}\right)^{2}\right)}{\left( 2a^{2}\left( X^{1}\right) ^{2}-m^{2}\mathcal{R}^{2}\cos(2X^{2})+m^{2}\mathcal{R}^{2}\right) ^{2}}\right] \xi^{2}(t) = 0,
\notag
\end{eqnarray}
\begin{equation}
{}
\end{equation}
\begin{eqnarray}  \label{57*}
&& \frac{{\rm d}^2\xi^{2}(t)}{{\rm d}t^2}+\frac{2a^{2}m^{2}Y^{1}\sin (2X^{2})}{2a^{2}\left(X^{1}\right) ^{2}-m^{2}\mathcal{R}^{2}\cos (2X^{2})+m^{2}\mathcal{R}^{2}}\frac{{\rm d}\xi^{1}(t)}{{\rm d}t}
\notag \\
&-&\frac{2m^{2}\mathcal{R}^{2}Y^{2}\sin(2X^{2})}{-2a^{2}\left(X^{1}\right)^{2}+m^{2}\mathcal{R}^{2}\cos(2X^{2})-m^{2}\mathcal{R}^{2}}\frac{{\rm d}\xi^{2}(t)}{{\rm d}t}
\notag \\
&+& \frac{2m^{2}\mathcal{R}^{2}}{a^{2}\left(X^{1}\right)^{3}}\left[ -\frac{2a^{4}\left(X^{1}\right)^{4}\sin(2X^{2})\left(a^{2}\left(\left(Y^{1}\right)^{2}-1\right)
+\mathcal{R}^{2}\left(Y^{2}\right)^{2}\right)}{\left(2a^{2}\mathcal{R}\left(X^{1}\right)^{2} - m^{2}\mathcal{R}^{3}\cos(2X^{2}) + m^{2}\mathcal{R}^{3}\right)^{2}}
- \frac{a^4C^2}{\mathcal{R}^4}\cot (X^{2})\right] \xi^{1}(t)
\notag \\
&+& \Bigg[ \frac{a^{2}C^{2}\csc^{2}(X^{2})\left(2a^{2}\left(X^{1}\right)^{2}
- m^{2}\mathcal{R}^{2}\right)}{\mathcal{R}^4\left(X^{1}\right)^{2}}+2\left( 2a^{2}\left(X^{1}\right)^{2}+m^{2}\mathcal{R}^{2}\right)
\nonumber\\
&\times&\frac{ \left(a^{2}\left(\left(Y^{1}\right)^{2}-1\right)
+ \mathcal{R}^{2}\left(Y^{2}\right)^{2}\right) }{-2a^{2}\mathcal{R}^{2}\left(X^{1}\right)^{2}+m^{2}\mathcal{R}^{4}\cos(2X^{2}) - m^{2}\mathcal{R}^{4}}\nonumber\\
&+& \frac{8a^{2}\left(X^{1}\right)^{2}\left(a^{2}\left(X^{1}\right)^{2}
+ m^{2}\mathcal{R}^{2}\right) \left(a^{2}\left(\left(Y^{1}\right)^{2}-1\right)
+ \mathcal{R}^{2}\left(Y^{2}\right)^{2}\right)}{\left(2a^{2}\mathcal{R}\left(X^{1}\right)^{2}
- m^{2}\mathcal{R}^{3}\cos(2X^{2}) + m^{2}\mathcal{R}^{3}\right)^{2}}
- 3\frac{a^4C^2}{\mathcal{R}^4}\csc ^{4}(X^{2})\Bigg]\nonumber\\
&\times &\xi^{2}(t)=0.
\end{eqnarray}

By evaluating Eqs.~(\ref{56*}) and (\ref{57*}) at $X^{1}=\pm2mC $, $X^{2}=\pm \sin^{-1}\left(2aC/\mathcal{R}\right)$, $Y^{1}=0$, $Y^{2}=0$, ($C^2 < (1/4)\mathcal{R^2}/a^2$), we obtain the evolution
equations of the deviation vector near the critical points as
\begin{equation}  \label{xi^1}
\frac{{\rm d}^2\xi^{1}(t)}{{\rm d}t^2}+\frac{1}{4m^2C^2}\xi ^{1}(t)=0,
\end{equation}
\begin{equation} \label{xi^2}
\frac{{\rm d}^2\xi^{2}(t)}{{\rm d}t^2}+\left(\frac{1}{4C^2} - \frac{a^2}{\mathcal{R}^2}\right)\xi^{2}(t)=0,
\end{equation}
and the general solutions are given by
\begin{equation} \label{sol-1}
\xi^{1}(t)= 2Cm\dot{\xi}^{1}_{0}\sin\left(\frac{1}{2mC} \ t\right),
\end{equation}%
\begin{equation}
\xi ^{2}(t) = \frac{C\mathcal{R}\dot{\xi}^2_0}{\sqrt{4a^2C^2-\mathcal{R}^2}}\left[\exp\left(\frac{\sqrt{4a^2C^2-\mathcal{R}^2}}{2C\mathcal{R}} \ t\right) - \exp\left(-\frac{\sqrt{4a^2C^2-\mathcal{R}^2}}{2C\mathcal{R}} \ t\right)\right],
\end{equation}
or, equivalently,
\begin{equation}\label{egq}
\xi ^{2}(t) = \frac{2C\mathcal{R}\dot{\xi}^2_0}{\sqrt{4a^2C^2-\mathcal{R}^2}}\sinh \left(\frac{\sqrt{4a^2C^2-\mathcal{R}^2}}{2C\mathcal{R}} \ t\right) ,
\end{equation}
where we have used the initial conditions $\xi ^{1}(0)=0$, $\xi ^{2}(0)=0$, $\dot{\xi}^{1}(0)=\dot{\xi} _{0}^{1}$, and $\dot{\xi}^{2}(0)=\dot{\xi} _{0}^{2}$. By taking into account that the condition of Jacobi stability requires $4a^2C^2-\mathcal{R}^2<0$, we can rewrite Eq.~(\ref{egq}) as
\be
\xi ^{2}(t) = \frac{2C\mathcal{R}\dot{\xi}^2_0}{\sqrt{\mathcal{R}^2-4a^2C^2}}\sin \left(\frac{\sqrt{\mathcal{R}^2-4a^2C^2}}{2C\mathcal{R}} \ t\right) ,
\ee
which indicates the oscillatory behavior of all the components of the geodesic deviation vector.

By evaluating Eqs.~(\ref{56*}) and (\ref{57*}) at $X^{1}=\pm \sqrt{C/(\pm \mathcal{R} - aC)}(m\mathcal{R}/\sqrt{a})$, $X^{2} = \pi/2$, $Y^{1}=0$, $Y^{2}=0$, ($C^2 = (1/4)\mathcal{R^2}/a^2$), we obtain the evolution
equations of the deviation vector near the critical points as
\begin{equation}  \label{xi^1*}
\frac{{\rm d}^2\xi^{1}(t)}{{\rm d}t^2}+\frac{4a^2(\mp a + \mathcal{R})^2}{m^2\mathcal{R}^4}\xi^{1}(t)=0,
\end{equation}
\begin{equation} \label{xi^2*}
\frac{{\rm d}^2\xi^{2}(t)}{{\rm d}t^2} = 0,
\end{equation}
which have the general solutions
\begin{equation} \label{sol-1*}
\xi^{1}(t)= \frac{\dot{\xi}^{1}_{0}m\mathcal{R}^2}{2a(aC \mp \mathcal{R})}\sin\left(\frac{2a(aC \mp \mathcal{R})}{m\mathcal{R}^2} \ t\right),
\end{equation}%
\begin{equation} \label{sol-2*}
\xi^{2}(t) = \dot{\xi}_0^2 t.
\end{equation}
Comparing Eqs. (\ref{xi^2}) and (\ref{xi^2*}) again implies $C^2 = (1/4)\mathcal{R}^2/a^2$, and substituting $C = \pm (1/2)\mathcal{R}/a$ into Eq. (\ref{xi^1*}) yields Eq. (\ref{xi^1}), as required by consistency. It may also be verified that, in this limit, Eq. (\ref{sol-1*}) becomes equivalent to Eq. (\ref{sol-1}). Finally, it may be shown that, for $C \rightarrow \pm (1/2)\mathcal{R}/a$, Eq. (\ref{egq}) is equivalent to Eq. (\ref{sol-2*}).

\section{Discussion} \label{S7}

In the present paper, we have investigated the numerical solutions of two wound string models, and have determined their Jacobi stability properties using Kosambi-Cartan-Chern (KCC) theory. The KCC theory introduces a geometric description of the time evolution of two-dimensional dynamical systems, described mathematically by second order differential equations, in which the solution curves are described by analogy with the theory of geodesics in a Finsler space. Using this approach, we have been able to describe the string dynamics in purely geometric terms, by means of a non-linear connection and the associated covariant derivative, while the stability properties of the system were obtained from the curvature deviation tensor.

In order to determine the Jacobi stability (or instability) of the string, we have estimated the eigenvalues of the deviation curvature tensor at the critical points of the system: if the real parts of the eigenvalues are strictly negative, the trajectories of the system bunch together at the critical point, whereas otherwise they diverge. Thus, if the trajectories near the critical point bunch together, the system is said to be Jacobi stable, whereas, if the diverge, the system is Jacobi unstable.

In our first model (Model I), we considered circular string loops, in $(3+1)$-dimensions, with windings in a higher-dimensional space approximated by a small patch of $\mathbb{R}^2$. This serves as a valid approximation to any compact internal manifold when the winding radius is small compared to the size of the compact dimensions, so that the nontrivial geometry and topology of the internal space can be neglected. In our second model (Model II), we considered string loops with windings in an $S^2$ internal space.

Our numerical solutions indicate that the dynamics of the string in the compact space (in both models) are highly sensitive to the initial conditions and the values of the model parameters, including the winding number, the initial loop size, and the initial velocities of the string in both the compact and non-compact directions. Furthermore, for the same initial conditions and choice of parameters, the higher-dimensional string dynamics differ greatly between Model I and Model II, therefore indicating (as expected) that the curvature and topology of the internal manifold play a nontrivial role in determining the higher-dimensional motion. However, we found that the motion of the string in the macroscopic dimensions was, in general, \emph{remarkably insensitive} to the {\it details} of the string motion in the higher dimensions, both within and ``between" models. That is, different sets of initial conditions in Model I (or Model II) gave rise to radically different motion in the compact space, but virtually identical motion in the macroscopic dimensions. Likewise, using the same set of initial conditions to solve the EOM for each model, we obtained very different types of motion in the higher dimensions, but negligible differences between the dynamics of the string in Minkowski space.

Generally, the periodicity of the loop radius in the Minkowski directions remained regular over large time scales, and depended on the higher-dimensional motion of the string only through the parameter $\Omega_0^2 \in (0,1)$, which represents the initial fraction of the total string length lying in the large dimensions. Though the macroscopic motion of the string \emph{is} sensitive to the value of $\Omega_0^2$, it may be shown that the motion of the higher-dimensional windings gives rise to an effective current from a $(3+1)$-dimensional perspective \cite{Nielsen:1979zf,Nielsen:1987fy}, such that $j(t_0) \propto (1-\Omega_0^2)/\Omega_0^2$ \cite{Yamauchi:2014ita,LaYo12}.  The key point is that the periodicity of the motion of the string in Minkowski space is \emph{unaffected} by how rapidly it oscillates in the internal manifold, even if the effective winding radius is also time-dependent. This suggests that many, qualitatively different, forms of higher-dimensional motion are degenerate from a $(3+1)$-dimensional perspective, giving rise to the same effective current under dimensional reduction. \emph{Therefore, our results indicate that, even if higher-dimensional strings exist, it is unlikely that the higher-dimensional nature of the string motion will leave any significant imprint on observable signatures in $(3+1)$ dimensions.}

In drawing these conclusions, it must be noted that the analysis performed here is purely classical, whereas quantum effects are likely to become significant for the motion of the string in the compact extra dimensions. Nonetheless, the equivalence of the higher-dimensional motion and the effective $(3+1)$-dimensional string current \cite{Nielsen:1979zf,Nielsen:1987fy} implies that classical treatment of the former is equivalent to classical treatment of the latter. Furthermore, the quantum theory of superconducting strings implies that such effects only become significant close to the threshold current $j_{\max} \sim q mc^2/\hbar$, where $q$ and $m$ denote the charge and mass of a fundamental charge carrier, respectively \cite{Witten85}. In the higher-dimensional model, $m \sim \hbar/(R c)$, where $R$ is the effective $(3+1)$-dimensional string width, which, in turn, is comparable to the effective winding radius \cite{Yamauchi:2014ita,LaYo12}. For bosonic currents, exceeding this threshold implies electric field strengths large enough to induce pair production via the Schwinger process, whereas, for fermionic currents, $j_{\max}$ marks the critical point at which it becomes energetically favorable for particles to leave the string \cite{Witten85}. In either case, $j = j_{\max}$ also marks the transition between strings with effective tension and strings with effective {\it pressure} in the {\it classical} model \cite{Copeland:1987th,Copeland:1987yv}. This causes loops to {\it expand} from their initial radius (before contracting), instead of first contracting (before re-expanding). In the higher-dimensional wound string model, this occurs when $\omega^2 >(<) 1/2$, respectively \cite{Yamauchi:2014ita,LaYo12}. Thus, in the critical case, $\omega^2 = 1/2$, which is equivalent to $j = j_{\max}$, the string is effectively tensionless (classically), from a $(3+1)$-dimensional perspective, and quantum instabilities may cause it to decay. However, for $\omega^2 \ll 1/2$, which corresponds to $j \ll j_{\max}$, the equivalence of higher-dimensional $F$-strings and $(3+1)$-dimensional superconducting strings, implies that the classical analysis remains valid.

The regularity of the periodic motion of the string in Minkowski space, obtained from the numerical solutions of the loop EOM, is consistent with the analytic results obtained via the KCC stability analysis. We found that, both in Model I and Model II, \emph{all} critical points of the wound string system were Jacobi stable.  Specifically, we considered the behavior of the geodesic deviation equations, describing the time variation of the deviation vector $\vec{\xi }$ near the critical points of the string system. For both models, the deviation equations take a simple form, and their solutions, giving the components of the deviation vector,  can be explicitly obtained. With one exception, the time-dependence of all components of $\vec{\xi }$ give rise to oscillatory behavior, corresponding to some bounded and finite values of the initial deviations. The exceptional case  occurs in Model II, when the string wraps great circles in the $S^2$, corresponding to the maximum possible winding radius. In this scenario, one component of the deviation tensor evolves linear in time, but the conditions for Jacobi stability are still satisfied. Thus, the explicit analysis of the deviation equations is in full agreement with our evaluation of the Jacobi stability conditions, and with our previous numerical results.

As proposed in \cite{Sa05}, the Jacobi stability of a system of second order differential equations can be interpreted as representing the {\it robustness} of a trajectory of the differential system, with respect to small perturbations of the {\it whole trajectory}. Therefore the mathematical concept of Jacobi stability is a very convenient and efficient way to describe the ``resistance" of limit cycles to small perturbations of the solutions of systems of differential equations. On the other hand, for systems of ordinary strongly non-linear differential equations,  we may give an alternative interpretation of the concept of Jacobi stability, as indicating the ``resistance" of a whole trajectory of the dynamical system to the onset of chaos. From both mathematical and physical points of view, the chaotic behavior in dynamical systems is assumed to develop due to small perturbations of the trajectory, and this interpretation is supported by one of the standard definitions of chaos. According to this definition, chaotic behavior occurs in systems of differential equations defined on a compact manifold $M$ once the geodesic trajectories grow exponentially fast, thus determining an exponential growth of the initial small deviations. Since chaotic behavior is closely related to the curvature of the base manifold $M$, or, more exactly, to the deviation curvature tensor $P_i^j$, the appearance of chaos in dynamical systems described by strongly nonlinear differential equations may be related to their Jacobi instability.

Moreover, it is important to point out that the concept of Jacobi (in)stability represents a natural generalization of the (in)stability of the geodesic flow on a metric differentiable manifold (Riemannian or Finslerian), which are extended to the general {\it non-metric} setting. Consequently, we may {\it conjecture} that Jacobi unstable trajectories of a non-linear system of differential equations could behave chaotically. From a physical point of view, this would imply that, after a finite interval of time, it would become possible to distinguish trajectories which, at the initial time $t_0$, were very near to each other. Alternatively, chaotic behavior actually implies sensitivity to initial conditions in the evolution of a dynamical system. Thus, a small change in the initial conditions could lead to a dramatic modification of the overall dynamics. More exactly, the sensitivity with respect to the  initial conditions means that, if we start with only a finite amount of information about the dynamical system (as is usually the case in most of the realistic applications), then, after a certain time, the behavior of the system will no longer be predictable.

Thus, by analyzing the issue of Jacobi stability for wound string models, we have analyzed the possibility of chaotic behavior developing in such systems. Physically, we would expect the onset of chaotic evolution to first become apparent in the microscopic motion of the string in the compact space, which varies over much shorter time and distance scales than its motion in the large dimensions. Nonetheless, once an instability develops, the highly nonlinear nature of the string EOM imply that it should, eventually, lead to chaotic evolution in $(3+1)$ dimensions. In this sense, the strong nonlinearilty of the higher-dimensional string equations gives rise to the possibility of chaotic $(3+1)$-dimensional motion, \emph{induced} by the onset of chaotic behavior in the full, $(3+n+1)$-dimensional space-time. However, our results support the initial {\it tentative} conclusion that the onset of chaos, at least in certain highly symmetric string systems (in this case, circular string loops) is, at best, an extremely late-time phenomenon. As such, it is unlikely to be observed, even if higher-dimensional strings exist, and even if their existence is one day experimentally confirmed.

\section*{Acknowledgments}

We would like to thank to the anonymous referee for her/his careful reading of our manuscript, and for comments and suggestions that helped us to improve it. We thank the Yat Sen School, the Department of Physics, and the Centre Sino-Fran{\c c}ais at Sun Yat Sen University for gracious hospitality during the final preparation of the manuscript. ML is supported by a Naresuan University Research Fund Individual Research Grant.



\begin{thebibliography}{99}

\bibitem{Frankel:1997ec}
  T. Frankel,
  {\it The Geometry of Physics: An Introduction},
  (Cambridge University Press, Cambridge, 1997)

\bibitem{Nakahara:1990th}
  M. Nakahara,
  {\it Geometry, topology and physics},
  Graduate student series in physics (Hilger, Bristol, 1990)

\bibitem{Pet1}
  M. Pettini,
  {\it Geometrical hints for a nonperturbative approach to Hamiltonian dynamics},
  Phys. Rev. \textbf{E 47}, 828 (1993)

\bibitem{Kau}
  H.E. Kandrup,
  {\it Geometric interpretation of chaos in two-dimensional Hamiltonian systems}, 
  Phys. Rev. \textbf{E 56}, 2722 (1997)

\bibitem{Pet0}
  M. Di Bari, D. Boccaletti, P. Cipriani, and G. Pucacco,
  {\it Dynamical behavior of Lagrangian systems on Finsler manifolds}, 
  Phys. Rev. \textbf{E 55}, 6448 (1997)

\bibitem{Pet2}
  P. Cipriani and M. Di Bari,
  {\it Finsler geometric local indicator of chaos for single orbits in the h{\' e}non-heiles hamiltonian}, 
  Phys. Rev. Lett. \textbf{81}, 5532 (1998)

\bibitem{Pet3}
  M. Di Bari and P. Cipriani,
  {\it Geometry and chaos on Riemann and Finsler manifolds}, 
  Planet. Space. Science \textbf{46}, 1543 (1998)

\bibitem{PR}
  L. Casetti, M. Pettini, and E.G.D. Cohen,
  {\it Geometric approach to Hamiltonian dynamics and statistical mechanics}, 
  Phys. Rep. \textbf{337}, 237 (2000)

\bibitem{Pet4}
  G. Ciraolo and M. Pettini,
  {\it Geometry of chaos in models of stellar dynamics}, 
  Celes. Mech. Dyn. Astron. \textbf{83}, 171 (2002)

\bibitem{Ko33}
  D.D. Kosambi,
  {\it Parallelism and path-spaces}, 
  Math. Z. \textbf{37}, 608 (1933)

\bibitem{Ca33}
  E. Cartan,
  {\it Observations sur le m{\' e}moire pr{\' e}c{\' e}dent},
  Math. Z. \textbf{37}, 619 (1933)

\bibitem{Ch39}
  S.-s. Chern,
  {\it  Sur la g{\' e}om{\' e}trie d'un systeme d'equations diff{\' e}rentialles du second ordre},
  Bulletin des Sciences Mathematiques \textbf{63}, 206 (1939)

\bibitem{An00}
  P. L. Antonelli (ed.),
  {\it Handbook of Finsler geometry}, 
  vol. 1 (Kluwer Academic, Dordrecht, 2003)

\bibitem{Mo}
  X.-h. Mo,
  {\it An introduction to Finsler geometry}, 
  Peking University Series in Mathematics, vol. 1 (World Scientific, Singapore, 2006)

\bibitem{ChernAMS}
  S.-s. Chern,
  {\it Finsler Geometry Is Just Riemannian Geometry without the Quadratic Restriction}, http://www.ams.org/notices/199609/chern.pdf

\bibitem{rev}
  C.G. Boehmer, T. Harko, and S. V. Sabau,
  {\it Jacobi stability analysis of dynamical systemsÑapplications in gravitation and cosmology},
  Adv. Theor. Math. Phys. 16, 1145-1196 (2012)

\bibitem{Sa05}
  S.V. Sabau,
  {\it Some remarks on Jacobi stability},
  Nonlinear Anal. \textbf{63}, 143 (2005)

\bibitem{Sa05a}
  S.V. Sabau,
  {\it Systems biology and deviation curvature tensor}, 
  Nonlinear Anal. Real World Appl. \textbf{6}, 563 (2005)

\bibitem{An93}
  P.L. Antonelli, 
  Tensor, N. S. {bf 52}, 27 (1993)

\bibitem{MHSS}
  R. Miron, D. Hrimiuc, H. Shimada, V.S. Sabau,
  \textit{The Geometry of Hamilton and Lagrange Spaces} 
  (Kluwer Acad. Publ., Dordrecht, 2001)

\bibitem{YaNa07}
  T. Yajima and H. Nagahama,
  {\it KCC-theory and geometry of the Rikitake system},
  J. Phys. A Math. Theor. \textbf{40}, 2755 (2007)

\bibitem{Ha1}
  T. Harko and V.S. Sabau,
  {\it Jacobi stability of the vacuum in the static spherically symmetric brane world models},
  Phys. Rev. D \textbf{77}, 104009 (2008)

\bibitem{Ha2}
  C.G. Boehmer and T. Harko,
  {\it Nonlinear stability analysis of the EmdenÐFowler equation},
  J. Nonlinear Math. Phys. \textbf{17}, 503 (2010)

\bibitem{T0}
  T. Yajima and H. Nagahama,
  {\it Nonlinear dynamical systems and KCC-theory},
  Acta Mathematica Academiae Paedagogicae Ny\'{\i}regyh\'{a}ziensis \textbf{24}, 179 (2008)

\bibitem{Punzi}
  R. Punzi and M. N. R. Wohlfarth,
  {\it Geometry and stability of dynamical systems},
  Phys. Rev. \textbf{E 79}, 046606 (2009)

\bibitem{T1}
  T. Yajima and H. Nagahama,
  {\it Tangent bundle viewpoint of the Lorenz system and its chaotic behavior},
  Phys. Lett. A \textbf{374}, 1315 (2010)

\bibitem{Ab1}
  H. Abolghasem,
  {\it Liapunov stability versus Jacobi stability},
  J. Dyn. Syst. Geom. Theor. \textbf{10}, 13 (2012)

\bibitem{Ab2}
  H. Abolghasem,
  {\it Jacobi stability of circular orbits in a central force},
  J. Dyn. Syst. Geom. Theor. \textbf{10}, 197 (2012)

\bibitem{Ab3}
  H. Abolghasem,
  {\it Stability of circular orbits in Schwarzschild spacetime},
  Int. J. Differ. Equ. Appl. \textbf{12}, 131 (2013)

\bibitem{Ab4}
  H. Abolghasem,
  {\it Jacobi stability of hamiltonian systems},
  Int. J. Pure Appl. Math. \textbf{87}, 181 (2013)

\bibitem{Ha3}
  T. Harko, C.Y. Ho, C.S. Leung, and S. Yip,
  {\it Jacobi stability analysis of the Lorenz system},
  Int. J. Geom. Methods Modern Phys. \textbf{12}, 1550081 (2015)

\bibitem{Ha4}
  T. Harko, P. Pantaragphong, and S.V. Sabau, 
  {\it A new perspective on the Kosambi-Cartan-Chern theory, and its applications},
  arXiv:1509.00168 (2015)

\bibitem{Ha5}
  T. Harko, P. Pantaragphong, and S.V. Sabau,
  {\it Kosambi-Cartan-Chern (KCC) theory for higher-order dynamical systems},
  Int. J. Geom. Methods Modern Phys. {\bf 13}, 1650014 (2016)

\bibitem{Zwi09}
  B.~Zwiebach,
  {\it A First Course in String Theory}, 
  2nd edn. (Cambridge University Press, Cambridge, 2009)

\bibitem{Polchinski:1998rq}
  J. Polchinski,
  {\it String Theory. Vol. 1: An Introduction to the Bosonic String} 
  (Cambridge University Press, Cambridge, 1998)

\bibitem{Polchinski:1998rr}
  J. Polchinski,
  {\it String Theory. Vol. 2: Superstring Theory and Beyond} 
  (Cambridge University Press, Cambridge, 1998)

\bibitem{Green:2012oqa}
  M.B. Green, J.H. Schwarz and E. Witten,
  {\it Superstring Theory, 25th Anniversary Edition} 
  (Cambridge University Press, Cambridge, 2012).

\bibitem{ViSh00}
  A. Vilenkin and E.P.S. Shellard,
  {\it Cosmic strings and other topological defects}.
  Cambridge Monographs in Mathematical Physics (Cambridge University Press, Cambridge, 2000).

\bibitem{And02}
  M.R. Anderson,
  {\it The Mathematical Theory of Cosmic Strings: Cosmic Strings in the Wire Approximation},
  (Taylor and Francis, Routledge, 2002).

\bibitem{Nielsen:1979zf}
  N.K. Nielsen,
  {\it Dimensional Reduction and Classical Strings},
  Nucl. Phys. B {\bf 167}, 249 (1980).

\bibitem{Nielsen:1987fy}
  N.K. Nielsen and P. Olesen,
  {\it Dynamical Properties of Superconducting Cosmic Strings},
  Nucl. Phys. B {\bf 291}, 829 (1987).

\bibitem{Yamauchi:2014ita}
  D.~Yamauchi and M.~J.~Lake,
  {\it Dynamics of cosmic strings with higher-dimensional windings},
  JCAP {\bf 1506} (06), 023 (2015)
  doi:10.1088/1475-7516/2015/06/023
  arXiv:1410.6267 [hep-ph]

\bibitem{Witten85}
  E. Witten,
  {\it Superconducting strings},
  Nucl. Phys. B {\bf 249},  557 (1985)


\bibitem{Go71}
  T. Goto
  {\it Relativistic quantum mechanics of a one-dimensional mechanical continuum and subsidiary condition of dual resonance model},
  Prog. Theor. Phys. {\bf 46}, 1560 (1971)

\bibitem{Na77}
  Y. Nambu
  {\it String-like configurations in Weinberg-Salam theory}
  Nucl. Phys. {\bf B130}, 505 (1977)

\bibitem{Dir75}
  P.A.M. Dirac,
  {\it The General Theory of Relativity} 
  (John Wiley and Sons, New York, 1975)

\bibitem{Denef:2007pq}
  F. Denef, M.R. Douglas and S. Kachru,
  {\it Physics of String Flux Compactifications},
  Ann. Rev. Nucl. Part. Sci. {\bf 57}, 119 (2007)
  arXiv: hep-th/0701050

\bibitem{LaWa10}
  M.~Lake, S.~Thomas and J.~Ward,
  {\it Non-topological Cycloops},
  JCAP {\bf 1001}, 026 (2010)
  doi:10.1088/1475-7516/2010/01/026 
  arXiv:0911.3118 [hep-th]; 
  M.~J.~Lake,
  {\it Cosmic necklaces in string theory and field theory},
  Ph.D. Thesis, Queen Mary, University of London (2010)

\bibitem{LaYo12}
  M.~Lake and J.~Yokoyama,
  {\it Cosmic strings with twisted magnetic flux lines and wound-strings in extra dimensions},
  JCAP {\bf 1209}, 030 (2012)
  Erratum: [JCAP {\bf 1308}, E01 (2013)]
  doi:10.1088/1475-7516/2013/08/E01, 10.1088/1475-7516/2012/09/030
  arXiv:1207.4891 [gr-qc]

\bibitem{S0}
  P.L. Antonelli and I. Buc\u{a}taru,
  {\it New results about the geometric invariants in KCC-theory},
  An. St. Univ. ``Al. I. Cuza", Ia\c{s}i {\bf 47}, 405 (2001)

\bibitem{S1}
  P.L. Antonelli, R.S. Ingarden and M. Matsumoto,
  {\it The theory of Sprays and Finsler Spaces with Applications in Physics and Biology} 
  (Kluwer Academic Publishers, Dordrecht, 1993)

\bibitem{S2}
  I. Buc\u{a}taru, O. Constantinescu, and M. Dahl,
  {\it A geometric setting for systems of ordinary differential equations},
  Int. J. Geom. Methods Mod. Phys. {\bf 8},  1291 (2011).

\bibitem{MiFr05}
  R. Miron and C. Frigioiu, 
  {\it On the Finslerian mechanical systems}, 
  Algebras Groups Geom. \textbf{22}, 151 (2005)

\bibitem{RH}
  Q. I. Rahman and G. Schmeisser,
  {\it Analytic Theory of Polynomials},
  London Mathematical Society Monographs. New Series 26 (Oxford University Press, Oxford, 2002)

\bibitem{Sarangi:2002yt}
  S. Sarangi and S.H.H. Tye,
  {\it Cosmic string production towards the end of brane inflation},
  Phys. Lett. B {\bf 536}, 185 (2002).
  arXiv:hep-th/0204074

\bibitem{Jones:2003da}
  N.T. Jones, H. Stoica and S.H.H. Tye,
  {\it The Production, spectrum and evolution of cosmic strings in brane inflation},
  Phys. Lett. B {\bf 563}, 6 (2003).
  arXiv:hep-th/0303269

\bibitem{Pogosian:2003mz}
  L. Pogosian, S.H.H. Tye, I. Wasserman and M. Wyman,
  {\it Observational constraints on cosmic string production during brane inflation},
  Phys. Rev. D {\bf 68}, 023506 (2003)
  [Phys.\ Rev.\ D {\bf 73}, 089904 (2006)].
  arXiv:hep-th/0304188

\bibitem{Copeland:1987th}
  E.J. Copeland, N. Turok and M. Hindmarsh,
  {\it Dynamics of Superconducting Cosmic Strings},
  Phys. Rev. Lett. {\bf 58}, 1910 (1987)

\bibitem{Copeland:1987yv}
  E.J. Copeland, D. Haws, M. Hindmarsh and N. Turok,
  {\it Dynamics of and Radiation From Superconducting Strings and Springs},
  Nucl. Phys. B {\bf 306}, 908 (1988)


\end{thebibliography}
\end{document}